\documentclass{aa}  

\usepackage{graphicx}
\usepackage{txfonts}
\usepackage{xcolor}
\usepackage{mathrsfs}
\usepackage[colorlinks=true, citecolor=blue]{hyperref}
\usepackage{physics}
\usepackage{nicefrac}
\usepackage[normalem]{ulem}
\usepackage{soul}
\usepackage[colorlinks=true, citecolor=blue]{hyperref}
\newcommand{\rossbi}{\textit{RoSSBi3D}}
\newcommand{\vr}{\vec{r}}
\newcommand{\el}{\vec{e}_l}
\newcommand{\es}{\vec{e}_s}
\newcommand{\eg}{\epsilon_g}
\newcommand{\ed}{\epsilon_d}
\newcommand{\ep}{\epsilon_p}
\newcommand{\edg}{\epsilon_{dg}}
\newcommand{\egc}{\epsilon_g(d_g)}
\newcommand{\Nabla}{\vec{\nabla}}
\newcommand{\fourier}{\mathcal{F}}
\newcommand{\er}{\Vec{e}_r}
\newcommand{\et}{\Vec{e}_\theta}
\newcommand{\Ldg}{L_{sg}^{dg}}
\newcommand{\Lg}{L_{sg}^{g}}
\newcommand{\Ld}{L_{sg}^{d}}
\newcommand{\Le}{L_{\epsilon}}

\newcommand{\isEquivTo}[1]{\underset{#1}{\sim}}

\begin{document}

\title{Self-gravity in thin-disc simulations of protoplanetary discs:}
\subtitle{smoothing length rectified and generalised to bi-fluids}

\author{S. Rendon Restrepo \inst{1,2,3}
\thanks{\email{srendon@aip.de}}
\and
P. Barge \inst{1}
\thanks{\email{pierre.barge@lam.fr}}
}
\institute{Aix Marseille Univ, CNRS, CNES, LAM, Marseille, France
\and 
Aix Marseille Univ, CNRS, Centrale Marseille, IRPHE, Marseille, France
\and
Leibniz-Institut für Astrophysik Potsdam (AIP), Potsdam, Germany}


 
\abstract
{To mimic protoplanetary discs (PPDs) evolution, 2D simulations with self-gravity must introduce a softening prescription of the gravitational potential. 
When the disc is only composed of gas the smoothing length is proportional to the gas scale height.
On the other hand when a dust component is included, the question arises as whether the smoothing length approach can still be used to quantify not only the dust self-gravity but also its gravitational interaction with gas.
}
{We identified grey areas in the standard smoothing length formalism for computing self-gravity in PPDs uniquely made of gas.
Our goal is to revisit the smoothing length approach which is then generalised to two phases when the dust component can be considered as a pressureless fluid.
}
{Analytical developments are used to approximate the vertically averaged self-gravity when the smoothing length is not assumed to be constant but rather a spatial function.
}
{
We obtained an analytical expression for the space varying smoothing length, which strongly improves the accuracy of the self-gravity computation.
For the first time, this method is generalised to address bi-fluid interactions in a PPD: two additional smoothing lengths are proposed for featuring an isolated dusty disc and gas-dust self-gravity interactions. 
On the computational ground, we prescribe the use of tapering functions for avoiding numerical divergences, checked that our method remains compatible with standard fast Fourier transform algorithms and evaluated computational costs. 
}
{Our space varying smoothing length permits (i) to solve the contradictions inherent to a constant smoothing length hypothesis, (ii) to fit accurately the 3D vertically averaged self-gravity and (iii) is applicable to a bi-fluid description of PPDs with the use of two additional smoothing lengths. 
Such results are crucial to enable realistic 2D numerical simulations accounting for self-gravity and are important to deepen our understanding of planetesimals formation and type I migration.
}


\maketitle
%

\section{Introduction}\label{sec: introduction}

Despite huge developments in 3D numerical computations and the advent of progressively sophisticated computational facilities, thin-disc (2D) simulations remain much less expensive and widely used in the study of protoplanetary discs (PPDs).
The 2D approximation lies on the vertical averaging of the 3D physical quantities and governing equations. 
When dealing with self-gravity (SG) the question becomes more sensitive since the equations cannot be vertically integrated.
In practise in PPDs studies an equivalent smooth potential, called a Plummer potential, is used to mimic as much as possible the vertically averaged SG force and, in this approximation, a smoothing length (SL) is introduced for accounting the disc vertical stratification.

As expected this approach applies to 2D studies of planet-disc interactions where different values have been suggested \citep{li_2009, dong_rafikov_2011}.
\citet{masset_2002,hure_pierens_2009} showed that, in the planet case, the SL should be proportional to the scale height of the gas disc and \citet{muller_kley_2012} proposed that $\epsilon_g/H_g=0.7$ where $\epsilon_g$ and $H_g$ are the gas SL and the pressure scale height, respectively.
\citet{muller_kley_2012} also explored the case of self-gravitating gas discs and found that, for vertically isothermal and stratified structures, one should instead use $\epsilon_g/H_g=1.2$ to avoid a systematic overestimate of the SG term.
They also showed that when accounting for vertical SG a deviation to the vertical Gaussian distribution occurs and in this case the SL is also proportional to the disc Toomre's parameter.

\citet{muller_kley_2012} prescription  clearly leads to small errors ($<2\%$) at large distances but it also has an important draw-back (never mentioned by the authors): SG is strongly underestimated at short distances with $100 \%$ errors !
Indeed, in agreement with \citet{hure_pierens_2009} the authors found that the accuracy in the approximation of SG terms is strongly improved if the SL is a space varying function \citep[Fig. 13]{muller_kley_2012}.
However, despite a discernible curve shape, and for unknown reasons, they didn't test analytical expressions that could fit at best the numerical curve. 
Further, they found that the best approximation is obtained when the SL vanishes at the singularity, a statement which is inconsistent with the divergence of the error in the SG computations (Figs. 12 and 13 of their paper). 
The reasons for this inconsistency are subtle mathematical details at the singularity that were unexplained by the authors and, at our knowledge, gone unnoticed until now.

In a bi-fluid description of PPDs, a dust layer is embedded in the gas disc and the evolution of gas and dust are coupled by aerodynamic forces. 
In this case, the momentum equations for the gas and the dust, both, contain two SG terms: one is due to the gas disc and the other to the dust layer. 
In other words, each fluid parcel (gas or dust) is submitted to the gravity of the gas disc and to the gravity of the dust sub-layer.
In contrast to the gas disc, the dust layer has a thickness which is governed not by pressure but turbulent stirring in the vertical direction. 
The scale height of the dust layer can be estimated in the form $H_d=\sqrt{\frac{\Tilde{\alpha}}{\Tilde{\alpha}+St}} H_g$ \citep{dubrulle_1995,weber_2019}, where $\Tilde{\alpha}=\alpha/{Sc}_z$, $\alpha$ is the dimensionless constant featuring turbulent $\alpha$-viscosity \citep{shakura_1973}, $Sc_z$ is the vertical Schmidt number and $St$ is the Stokes number.
Vertical averaging over the dust layer faces the same numerical issues than averaging over the gas disc. 
The problems can be solved in the same way than for a single gas disc and a simple extrapolation provides the dust disc SL: $\epsilon_d \propto 1.2 H_d $. 
Since the dust layer is much thinner than the gas disc (due to dust settling) the two SLs are very different from one another with $\epsilon_d << \epsilon_g$.
Thus, any error between these two parameters may lead to an incorrect and significant underestimation of the dust layer contribution to the SG terms.
The question becomes more complex when accounting for the crossed SG terms.
At present time we do not possess any rigorous theoretical approach permitting to evaluate this gravitational bi-fluid interaction.
But we think that this problem can also be addressed with a SL and, intuitively, we can expect that it should lie between the gas SL and the dust SL. 
To our knowledge, both issues raised in this paragraph for bi-fluids are new and were never addressed in numerical and theoretical studies of PPDs.

The improvements we performed are crucial to enable realistic 2D numerical simulations when SG is taken into account and are also particularly important in planet formation theories to better understand the formation of planetesimals. 
Indeed, it makes possible the study of the gravitational fragmentation of the dust layer \citep{1973_goldereich}, the formation of coherent clumps under the streaming instability \citep{johansen_youdin_2007}, the gravitational clumping of dust particles trapped in a large-scale vortex \citep{barge_1995} or in a co-orbital trapping scenario \citep{barge_2022}.
Indirectly, we found that this work could also have implications in the study of type I planet migration.

In this paper, our goal is to provide a method, based on the SL formalism, to accurately compute SG terms, in thin disks, at short and large separations. 
We aim to introduce the substantial SG interaction of gas and dust, when latter is considered as a pressureless fluid.
We begin in Sect. \ref{sec:A correct estimation of SG in bi-fluid simulations} conducting a theoretical development from first principles which justifies the use of the Plummer potential.
Then, in Sect. \ref{sec: mutual interactions} we explore SG estimation through the prism of the SL paradigm.
We corrected and completed the Plummer potential formalism thanks to a spatial dependent SL and generalise it to the case of bi-fluids. 
In Sect. \ref{sec: numerical treatment} we handle practical aspects such as numerical divergences, computational costs and the calculation of the corrected SG thanks to fast Fourier transforms (FFT) methods.
Finally, in Sect. \ref{sec:discussion} we suggest a discussion followed by a conclusion.

\section{Self-gravity terms for bi-fluid simulations}\label{sec:A correct estimation of SG in bi-fluid simulations}

In this Sect. we set up the theoretical background for computing gas and dust SG contributions, when solid material is considered as a pressureless fluid, in a thin-disc approximation.
In this context we also remind the interest of the Plummer potential for 2D SG calculations and generalise this formalism to bi-fluids.

\subsection{2D approximation and formal derivation}

In 3D, the SG force per unit volume exerted by the PPD on a gas and dust parcels are:
\begin{equation}\label{Eq:3D volume force total}
\begin{array}{lll}
\vec{f}^{g,tot}_{3D}(\vr,z) &=& \rho_g(\vr,z) \left[ \Nabla \Phi_g + \Nabla \Phi_d \right] \\ [4pt]
\vec{f}^{d,tot}_{3D}(\vr,z) &=& \rho_d(\vr,z) \left[ \Nabla \Phi_g + \Nabla \Phi_d \right] 
\end{array}
\end{equation}
where $\Phi_g$ and $\Phi_d$ are the distinct gravitational potentials of the gas and dust discs, respectively.
For the sake of generality and conciseness, we denote both fluid phases as \textbf{a} and \textbf{b} which reduces each term of the r.h.s of Eqs. \ref{Eq:3D volume force total} to:
\begin{equation}
\begin{array}{llll}
\vec{f}^{a \rightarrow b}_{3D}(\vr,z) 
                    &=& \rho_b(\vr,z) \Nabla \Phi_a                            \\ [6pt]
                    &=& \displaystyle- G \, \rho_b(\vr,z)                      \\
                    & & \displaystyle \iint\limits_{disc} \int\limits_{z'=-\infty}^{+\infty} 
                        \frac{ \rho_a(\vr',z')}{||\vr-\vr'||^2+(z-z')^2} \,
                        \el \, d^2\vr' \, dz' 
\end{array}
\end{equation}
where $\vec{f}^{a \rightarrow b}_{3D}$ is force per unit volume the \textbf{a}-disc exerts on an elementary \textbf{b} fluid element.
The density of phase \textbf{a} is noted $\rho_a$ and $\el=\left[ \, \vr-\vr' + (z-z') \, \vec{e}_z  \right] \big/ \sqrt{ ||\vr-\vr'||^2 + (z-z')^2 }$.
Assuming a vertical hydrostatic equilibrium and an isothermal approximation in the vertical direction for gas, the volume density can be written as $\rho_g(\vr, z)= \rho_{0,g}(\vr) \, e^{-\frac{1}{2} z^2/H_g(\vr)}$, where $H_g(\vr)$ is the gas pressure scale height.
We assume same vertical Gaussian profile for dust density than for gas but this time the vertical equilibrium is rather governed by turbulent stirring which sets a different scale height for dust: $H_d(\vr)$.
This permits to write both surface densities as:
\begin{equation}
\Sigma_a(\vr) = \int\limits_{z=-\infty}^{\infty} \rho_{a}(\vr,z) \, dz = \sqrt{2 \pi} \, H_a(\vr) \, \rho_{0,a}(\vr)
\end{equation}
The 2D analogue of the SG force is simply obtained integrating the 3D SG force in the vertical direction:
\begin{equation}
\begin{array}{lll}
\vec{f}^{a \rightarrow b}_{2D}(\vr) 
             & = & \int\limits_{z=-\infty}^{\infty} 
                   \vec{f}^{a \rightarrow b}_{3D}(\vr,z) \, dz \\
             & = & - G \, \rho_{0,b}(\vr) 
                   \iint\limits_{disc} 
                   \,\rho_{0,a}(\vr') \, s \, \es  \\
             &   & \displaystyle \left( \, \, \int\limits_{z, z'=-\infty}^{+\infty} 
                   \displaystyle \frac{e^{-\frac{1}{2} z^2/H_b^2(\vr)} e^{-\frac{1}{2} z'^2/H_a^2(\vr')}}{\left(s^2+(z-z')^2\right)^{{3}/{2}}} \, dz \, dz' \right) d^2\vr'
\end{array}
\end{equation}
where $s=||\vr-\vr'||$ is the separation (or mutual distance) between two fluid elements and $\es=\left(\vr-\vr'\right)\big/s$.
In this 2D approximation, it is implicitly assumed that the disc is symmetric with respect to the $z=0$ plane which allows to cancel naturally the vertical component of above force during the integration. 
After variables substitution, we finally obtain:
\begin{equation}\label{Eq: volume force 2D}
\vec{f}^{a \rightarrow b}_{2D}(\vr) 
             = - \frac{G}{\pi} \, \Sigma_{b}(\vr) 
               \iint\limits_{disc} 
               \, \frac{\Sigma_{a}(\vr')}{H_g(\vr) \, s} \, L^{a b}_{sg} (d_g, d_b, \eta)  \, \es \, d^2\vr'
\end{equation}
where:
\begin{equation}\label{Eq: SGFC general}
L^{a b}_{sg} (d_g, d_b, \eta_{ab}) 
            =  \displaystyle \frac{1}{2} \frac{d_b^3(\vr)}{d_g(\vr)} \iint\limits_{u,v=-\infty}^{+\infty} 
               \frac{e^{-\frac{u^2}{2}} e^{-\frac{v^2}{2}}}{\left[ d_b(\vr)^2+(u-\eta_{a b} v)^2 \right]^{3/2}}
               \, du \, dv
\end{equation}
is a normalised quantity that we called self-gravity force correction (SGFC)\footnote{\citet{muller_kley_2012} defined as force correction a quantity that they defined as $I_{sg}$.
We noticed that in some of their graphics they mistook $s I_{sg}(s)$ with $I_{sg}(s)$. 
Therefore for clarity and consistency with their work, we did not adopted their naming convention.} (with respect to the 3D case), $d_b(\vr) = s/H_b(\vr)$ is the separation normalised with respect to  \textbf{b} scale height and $\eta_{ab} = H_a(\vr')/H_b(\vr)$ is the \textbf{a}-to-\textbf{b} scale height ratio.
In the following we will use the notation $d_b=d_b(\vr)$, except when a distinction is necessary. 
We want to highlight that in our reasoning $\eta_{ab}$ is a spatial varying quantity, but for the rest of this article we will rather use:
\begin{equation}
\eta_{ab}=\langle H_a(\vr') \rangle / \langle H_b(\vr) \rangle
\end{equation}
where $\langle \rangle$ stands for space averaging over the 2D disc.
This simplifies next theoretical developments.
We note that the scale heights used in this paper could be time-varying functions, but for conciseness, the time dependence is not explicit in the equations.

The numerical determination of SG in thin-disc and bi-fluid simulations mainly relies on the computation of the SGFC, but no analytical expression of this integral has been found in terms of standard mathematical functions. 
This is why, a Plummer potential is commonly used to approximate at best this integral.

\subsection{Fitting self-gravity terms thanks to smoothing lengths}

Both vertical averaging in Eq. \ref{Eq: SGFC general} cannot be performed analytically but in practice we can introduce a Plummer potential, $\Psi^{ab}_\epsilon$, in the 2D gravitational potential, $\Phi^{ab}_\epsilon$, which permits to approach the 2D SG force defined in Eq. \ref{Eq: volume force 2D}:
\begin{equation}\label{Eq: Total SG force with epsilon}
\begin{array}{lll}
\vec{f}^{a \rightarrow b}_{2D, \epsilon}(\vr) 
             & = & \displaystyle - \frac{\Sigma_{b}(\vr)}{\pi} \, \Nabla_s \Phi^{ab}_{\epsilon}(\vr) \\ [6pt]
             & \mbox{with} & \displaystyle\Phi^{ab}_{\epsilon}(\vr)
               = -G \iint\limits_{disc} \Sigma_a(\vr') \, \Psi_{\epsilon}^{ab}(s) \, \dd^2 \vr'
\end{array}
\end{equation}
where:
\begin{equation}\label{Eq:plummer potential}
\Psi^{ab}_{\epsilon}(s)={\pi}\big/\left({s^2+\epsilon_{ab}^2}\right)^{1/2}
\end{equation}
and $\epsilon_{ab}$ is the SL between phases \textbf{a} and \textbf{b}.
Usually $\epsilon_{ab}$ is assumed to be constant, so that:
\begin{equation}\label{Eq:derivative plummer potential}
||\nabla \Psi^{ab}_{\epsilon}||= \pi s \big/\left({s^2+\epsilon_{ab}^2}\right)^{3/2}
\end{equation}
For a gas disc, \citet[Fig. 13]{muller_kley_2012} have shown that the SL that gives the best fits to the SGFC is a spatial function of $d_g$ that is, in fact, inconsistent with Eq. \ref{Eq:derivative plummer potential}.
Indeed, the additional term $\partial_s \epsilon_{ab}^2 /2$ should be present in the numerator of aforementioned equation.
So, in order to remain mathematically correct and to keep the possibility to make comparisons with the work of \citet{muller_kley_2012} we decided to consider the potential satisfying Eq. \ref{Eq:derivative plummer potential} and not Eq. \ref{Eq:plummer potential}.
This is a slight change in the SL paradigm that does not affect the approximation of the 2D SG terms and leads to meaningful 2D results.
Based on this clarification, we can define an analogue of the SFGC that is compatible with the SL approach:
\begin{equation}\label{Eq: force correction SL}
\begin{array}{ccc}
L_\epsilon^{a b} (d_g)
        & = & H_g(\vr) \, s \, ||\nabla \Psi_{\epsilon}^{ab}||  \\ [4pt]
        & = & \displaystyle \frac{\pi d_g^2}{\left[ d_g^2 + (\epsilon_{ab}(d_g)/H_g(\vr))^2 \right]^{3/2}}
\end{array}
\end{equation}
This normalised quantity, which we have called the smoothing length force correction (SLFC), should fit the SGFC to correctly estimate SG in 2D simulations.
This is only possible choosing wisely a spatially depending SL, $\epsilon_{ab}(d_g)$, as it will be depicted in next Sect.
The definition and abbreviation of the main quantities encountered in this paper are given in Table \ref{tab:list abbreviations}.

%
\begin{table}
\caption{Definitions and list of abbreviations}             
\label{tab:list abbreviations}      
\centering          
\begin{tabular}{l l l}     
\hline\hline       
Abbrev.             & Definition/Name                   & Symbol                              \\ 
\hline              
SG                  & Self-gravity                      &                                     \\ 
                    & Mutual distance or separation     & $s = ||\vr-\vr'||$                  \\
                    & Normalised separation for         & $d_b = s/H_b$                       \\
                    & phase b                           &                                     \\  
SL                  & Smoothing length                  &                                     \\
CSL                 & Constant smoothing length         & $\epsilon_{ab}=const.$              \\  
SVSL                & Space varying smoothing length    & $\epsilon_{ab}(d_g)$                \\
SGFC                & Self-gravity force correction     & $L^{a b}_{sg}$ (Eq. \ref{Eq: SGFC general})             \\
SLFC                & Smoothing length force correction & $L_\epsilon^{a b}$ (Eq. \ref{Eq: force correction SL})  \\
PDFC                & Planet-disc force correction      & $L^p$ (Eq. \ref{Eq: planet-disc force correction}) \\
                    & \textbf{a}-to-\textbf{b} scale height ratio & $\eta_{ab}$               \\
                    & Gas-to-dust scale height          & $\eta=\langle H_g\rangle /\langle H_d \rangle $ \\
\hline                  
\end{tabular}
\end{table}

\section{Mutual self-gravity interactions based on the smoothing length approach}\label{sec: mutual interactions}

We aim here to provide a suitable SL that most closely approaches each of the possible exact gravitational interactions: gas-gas, dust-dust and dust-gas.
We start by retrieving and rectifying \citet{muller_kley_2012} results for a disc only made of gas.

\subsection{Contribution of the gas disc on a gas parcel}\label{subsec:gas and gas}

\begin{figure}
\centering
\includegraphics[width=\hsize]{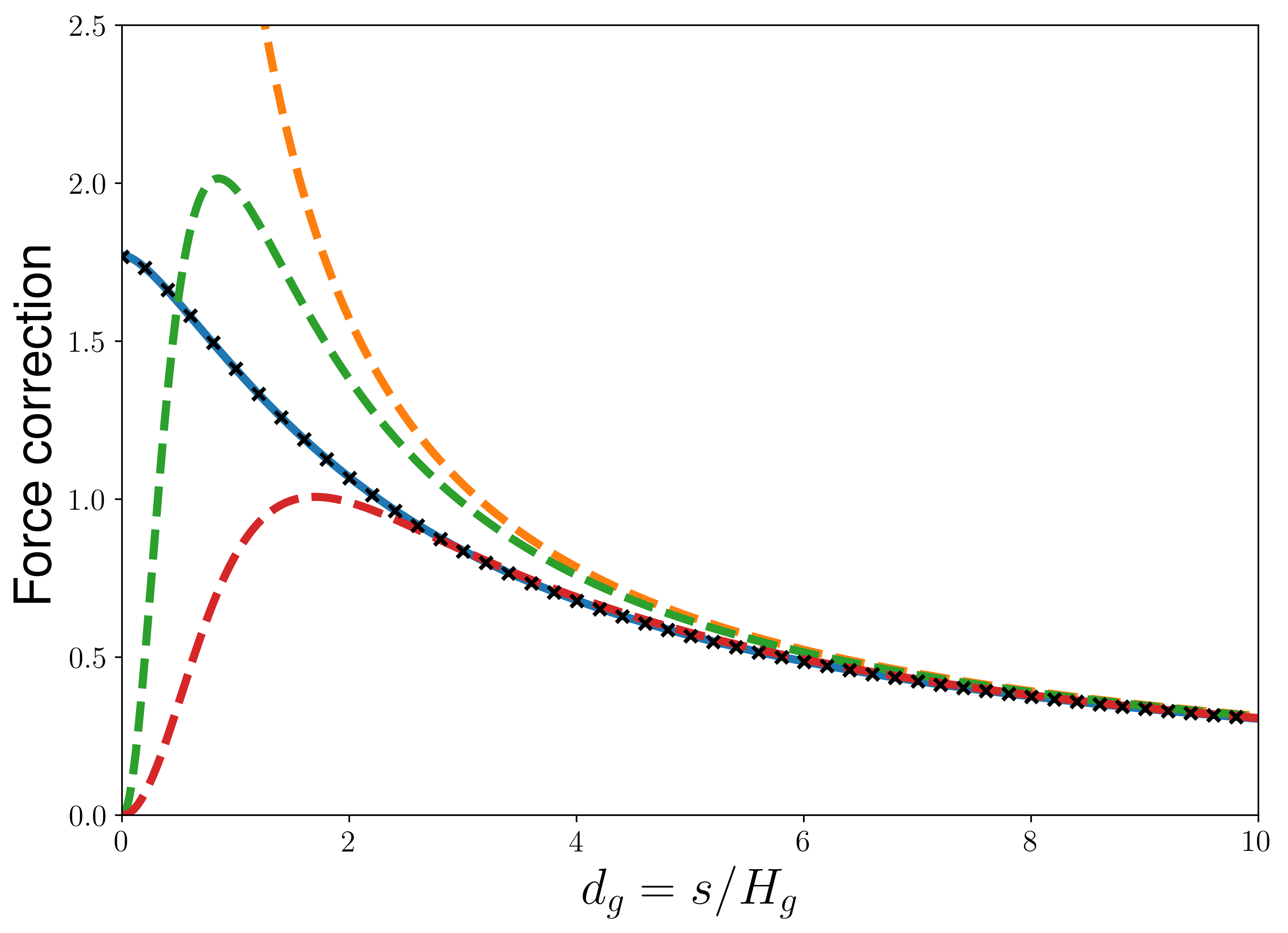}
\includegraphics[width=\hsize]{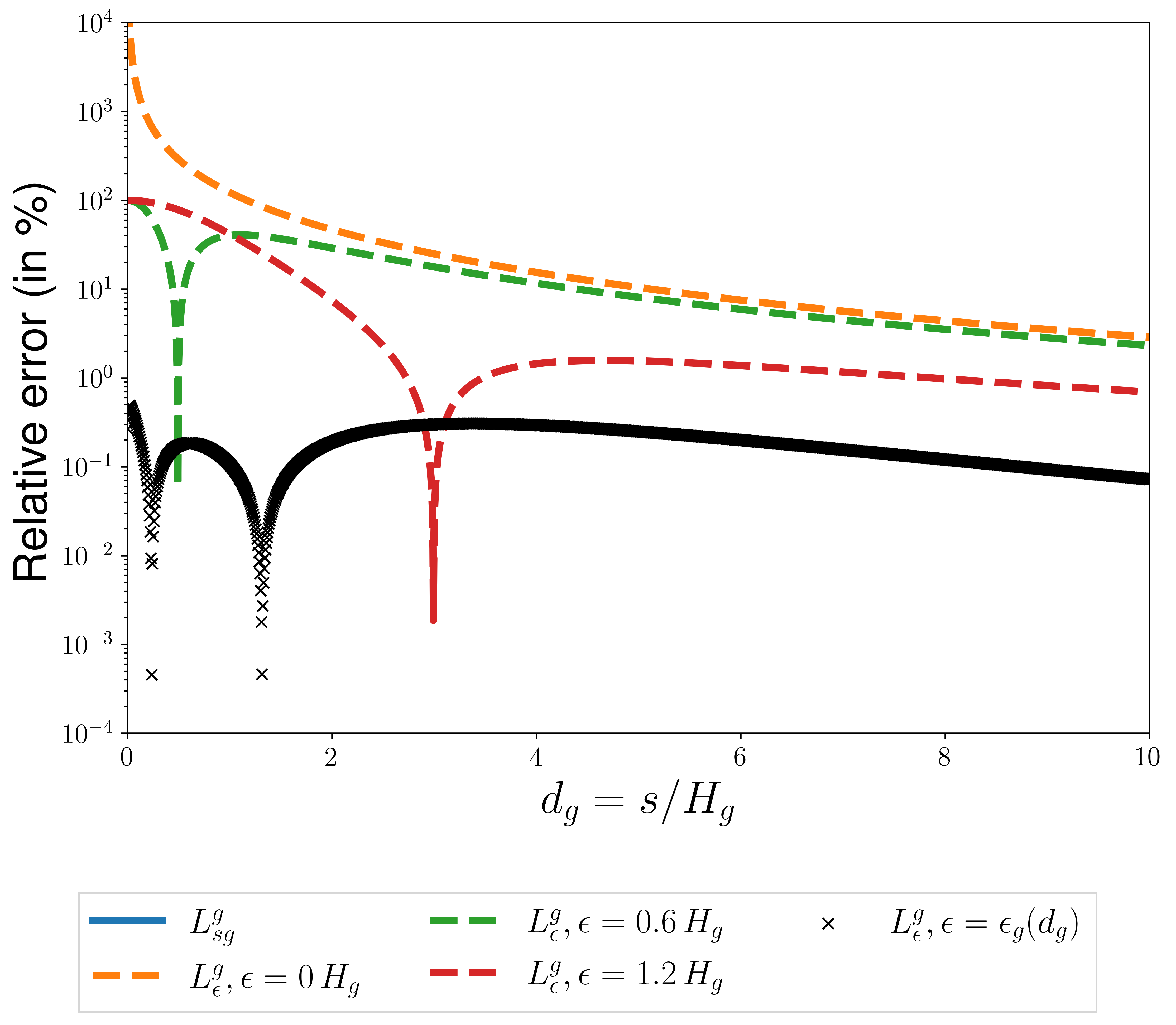}
\caption{\textbf{SG and SL force corrections for gas only.} \\
\emph{Top:} SGFC (exact value, in blue solid line) and SLFC for different smoothing lengths based on constant values for the SL (red, green and orange dashed lines)) or on function $\epsilon_g(d_g)$ (black cross markers).\\
\emph{Bottom:} Relative error  between SGFC and SLFC (in \%):
$100\cdot|\mbox{SLFC}/\mbox{SGFC}-1|$                       \\
For a CSL, $\epsilon_g=const.$, the SLFC and SGFC don't match at short separations: the error curve either tends either towards 0 or towards infinity. 
On the contrary, for a SVSL, $\epsilon_g=\epsilon_g(d_g)$, the SLFC matches SGFC with an accuracy better than 0.5\%. 
We found that $L_0=\sqrt{\pi}$ (Appendix \ref{app: lambda calculation}). } 
\label{fig: Lg and error_g}
\end{figure}

\begin{figure}
\centering
\includegraphics[width=\hsize]{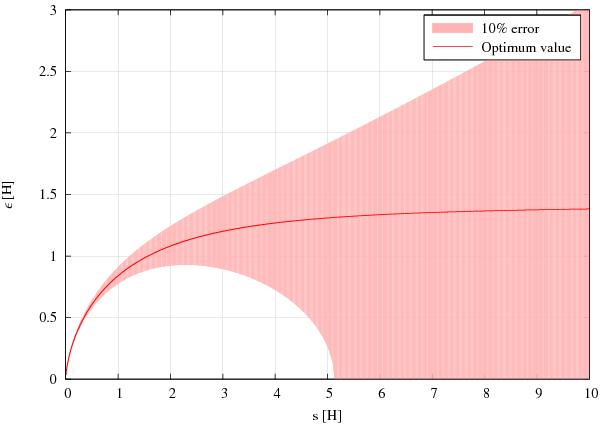}
\includegraphics[width=\hsize]{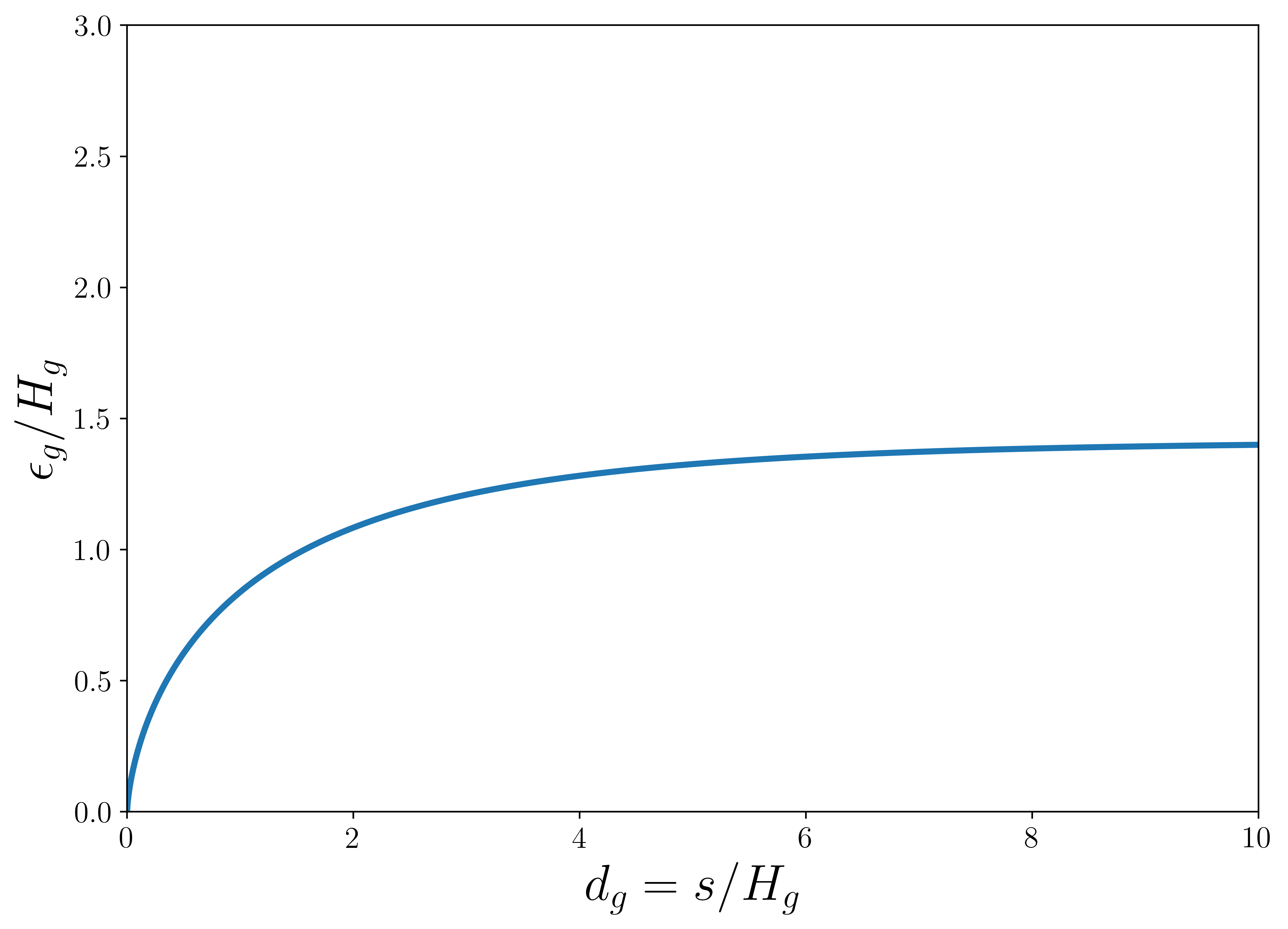}
\caption{\textbf{Space varying smoothing length for gas only}                    \\
\emph{Top:} SVSL obtained numerically by \citet{muller_kley_2012}.               \\
\emph{Bottom:} SVSL obtained in this work by analytical computations and curve fitting.}
\label{fig: SL for gas}
\end{figure}

This case corresponds to \textbf{a}=\textbf{b}=g.
From the general definition provided by Eq. \ref{Eq: SGFC general} the gas SGFC is equal to \footnote{For sake of simplicity in the whole article we replaced all superscripts $gg$ by $g$.}:
\begin{equation}\label{Eq: force correction gas-gas}
L_{sg}^{g}(d_g)
            = \displaystyle \frac{1}{2} d_g^2
              \iint\limits_{u,v=-\infty}^{\infty} 
              \frac{e^{-\frac{u^2}{2}} e^{-\frac{v^2}{2}}}{\left[ d_g^2+(u-\eta_g v)^2 \right]^{3/2}}
              \, du \, dv
\end{equation}
where $\eta_g=\langle H_g(\vr')\rangle / \langle H_g(\vr) \rangle=1$ \footnote{We think that \citet{muller_kley_2012} implicitly assumed this equality in their work.}.
We present in Fig. \ref{fig: Lg and error_g} top panel above exact SGFC (blue solid line), evaluated numerically, and the equivalent SLFC (dashed lines) for three constant smoothing lengths (CSLs): $\eg(d_g)/H_g(\vr)=[0.0, 0.6, 1.2]$.
In the bottom panel are showed the respective errors in percentage.
These curves testify the retrieval of \citet{muller_kley_2012} results in the case of a self-gravitating disc which validates our approach.
For long distances the three aforementioned SLFC have the same behaviour and the error becomes negligible.
For instance the CSL value $\egc=1.2 H_g$ offers the smallest error at long distances, less than $2\%$ for $d_g \geq 4$.
On the contrary, at short distances the SLFC either vanishes or diverges which leads to significant errors since the exact SGFC converges towards $L_0 \simeq 1.772$.
Such behaviour is indeed intrinsic to the CSL formalism at short distances:
\begin{equation}\label{Eq: limits constant smoothing length}
\begin{array}{cccll}
\mbox{if} & \eg(d_g)=0 & : &
             \displaystyle L_{\epsilon}(d_g) = \frac{\pi}{d_g} 
             & \rightarrow + \infty \\
\mbox{if} & \eg(d_g)=const. \neq 0 & : &
             \displaystyle L_{\epsilon}(d_g)  \isEquivTo{d_g \rightarrow 0} \pi \left(\frac{H_g}{\eg}\right)^3 d_g  
             & \rightarrow 0
\end{array}
\end{equation}
Above limits demonstrate that the CSL formalism either leads to 100 \% or infinite errors at short distances when $\eg=const.\neq0$ and $\eg=0$, respectively.   
This divergence between the SGFC and SLFC was not highlighted by \cite{muller_kley_2012} and a possible reason is that this behaviour is suitable since it allowed to avoid numerical divergences at $d_g=0$ when $\eg(d_g)=const. \neq 0$.
Even if \citet{muller_kley_2012} didn't shine a light on these divergences at short distances they nevertheless evaluated numerically that a space varying smoothing length (SVSL) was necessary to fit the exact SGFC, that we reproduced in top panel of Fig. \ref{fig: SL for gas}.
From this Fig. we can infer that the SVSL should tend to 0 in order that the SLFC matches the SGFC.
But this is in contradiction with Fig. \ref{fig: Lg and error_g} and Eqs. \ref{Eq: limits constant smoothing length}, since the SLFC is supposed to diverge analytically for $\epsilon_g(d_g)=0$ at short distances.
In general, no matter the chosen CSL, the gas SG was underestimated by a factor 100 at short distances.
This contradiction went unnoticed and we intend to solve it in next paragraph.

The analytic expression of the gas SL which better fits Fig. \ref{fig: SL for gas} top panel and thus permits the SLFC to fit the exact SGFC should fulfil next constraints:
\begin{equation}
\begin{array}{ccc}
\lim\limits_{d_g \rightarrow 0}       \egc & = & 0                    \\
\lim\limits_{d_g \rightarrow +\infty} \egc & = & \sqrt{2} \, H_g(\vr) \\
\Le^{g}(d_g=0)                                 & = & L_0 
\end{array}
\end{equation}
Thanks to analytical arguments completed by a curve fit, made explicit in appendix \ref{app: gas SL derivation}, we found a model for the SVSL which permits to accurately approach the SGFC:
\begin{equation}\label{Eq: SVSL}
\displaystyle \egc=\sqrt{2} \, H_g(\vr) \, \left[1-\exp\left(-\frac{\epsilon_{g,0}}{\sqrt{2}} d_g^{2/3}-\alpha d_g^n\right) \right] 
\end{equation}
where $\epsilon_{0,g}=\left[ {\pi}/{L_0} \right]^{1/3}$ and the numerical values of $(\alpha, n)$ are gathered in Table \ref{tab: fitting parameters}.
In particular, the power $2/3$ and $\epsilon_{0,g}$ allow to reach the finite value $L_0$ for the SLFC at $d_g=0$.
In Fig. \ref{fig: SL for gas} bottom panel we show the SVSL analytic model and in Fig. \ref{fig: Lg and error_g} the respective SLFC and error with respect to the exact SGFC (black cross markers).
Within this correction, the SLFC and SGFC overlap and the error is reduced to less than 0.5 \% in the whole distance range.
As a matter of comparison, this correction allows to decrease the error up to factors 200 and 40 at short and long distances, respectively, compared to the CSL where $\epsilon(d_g)=1.2 H_g(\vr)$.

\begin{table}
\caption{Fitting parameters for space varying smoothing lengths and $\delta$ models} 
\label{tab: fitting parameters}                            
\centering                                                 
\begin{tabular}{c | c}                                     
\hline                                                     
$(\alpha,n)$  & $(0.04319874, 1.14791757)$  \\  [6pt]     
\hline
$(\beta, q)$  & $(0.06427627, 1.14735482)$  \\  [6pt]
\hline
$(\gamma,m)$  & $(0.98362092, 0.75552227)$  \\  [6pt]
\hline
\end{tabular}
\end{table}

\subsection{Contribution of the dust layer on a dust parcel}\label{sec:dust and dust}

\begin{figure}
\centering
\includegraphics[width=\hsize]{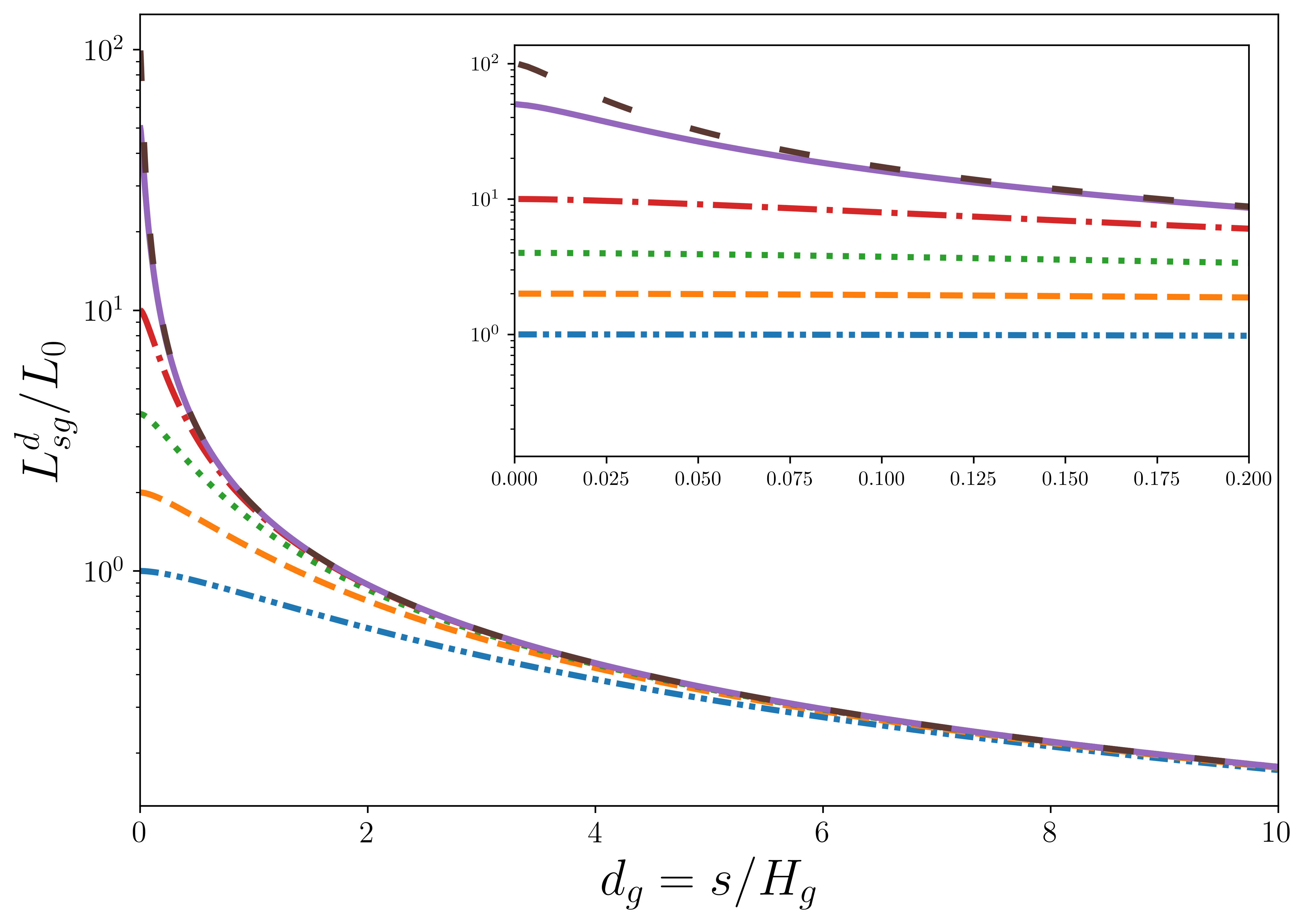}
\includegraphics[width=\hsize]{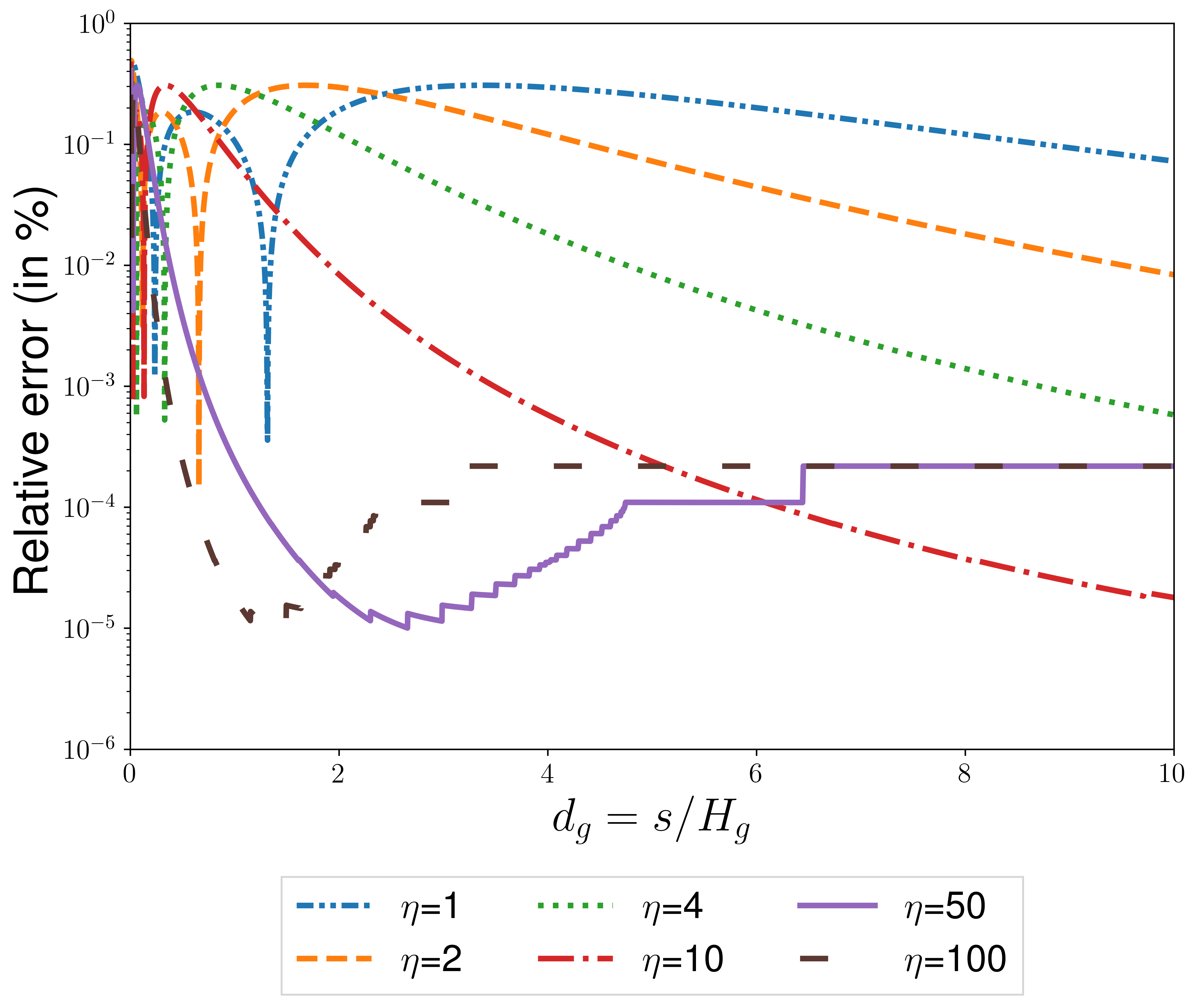}
\caption{\textbf{SG and SL force corrections for dust only}                                         \\
\emph{Top:} SGFC for different values of the gas-to-dust scale-height ratio ($\eta$)                   \\
\emph{Bottom:} Relative error between SGFC and SLFC (in \%): $100\cdot|\mbox{SLFC}/\mbox{SGFC}-1|$. \\
For the whole range of $\eta$ values the accuracy is better than 0.5 \%.
The dust SGFC is proportional to $\eta$ at short distances ($d_g\leq1.5/\eta$).
For large $\eta$, this could favour gravitational clumping.
}
\label{fig: Ld and error_d}
\end{figure}

In this Sect. we take \textbf{a}=\textbf{b}=d.
From the general definition provided by Eq. \ref{Eq: SGFC general} the dust SGFC is \footnote{For sake of simplicity in the whole article we replaced all superscripts $dd$ by $d$.}:
\begin{equation}\label{Eq: force correction dust-dust}
\begin{array}{ll}
\Ld(d_g, d_d)
         & = \displaystyle\frac{1}{2} \frac{d_d^3}{d_g} 
               \iint\limits_{u,v=-\infty}^{\infty}
               \frac{e^{-\frac{u^2}{2}}e^{-\frac{v^2}{2}}}{\left[d_d^2+(u-\eta_d v)^2\right]^{3/2}} \, du \, dv \\
         & = \displaystyle \frac{1}{2} \eta \left(\eta d_g \right)^2 
               \iint\limits_{u,v=-\infty}^{\infty}
               \frac{e^{-\frac{u^2}{2}}e^{-\frac{v^2}{2}}}{\left[(\eta d_g)^2+(u-v)^2\right]^{3/2}} \, du \, dv    \\
         & = \eta L_{sg}^{g} (\eta d_g)
\end{array}
\end{equation}
where $\eta_d=\langle H_d(\vr') \rangle /\langle H_d(\vr) \rangle=1$ and $\eta=H_g(\vr)/H_d(\vr)$ is the gas-to-dust scale height ratio that, in general, is greater than unity. 
Again for simplicity we rather use in the rest of this article $\eta=\langle H_g(\vr)\rangle /\langle H_d(\vr) \rangle $.
Above relation naturally applies to the dust SLFC but it is convenient to write it in two different ways:
\begin{equation}
L_{\epsilon}^{d}(d_g) 
            =  \eta L_{\epsilon}^{g} (\eta d_g) \quad \mbox{ or } \quad \ed(d_g, \eta) = \frac{\eg(\eta d_g)}{\eta}
\end{equation}
Both equations are equivalent but the former provides a physical insight while the latter allows a simple implementation in hydrodynamical codes.
In Fig. \ref{fig: Ld and error_d} top and bottom panels are shown the SLFC and the error with respect to the SGFC for different gas-to-dust scale height ratios, respectively.
The dust SLFC and SGFC curves overlap for the whole $\eta$ and distances range, so we did not plot these two quantities in the same figure to avoid duplication.
The error is again lower than 0.5 \% in the whole distance range as expected from the unique gas disc case.
We want to highlight that at long separations the dust SLFC matches the gas SLFC, $\Ld(d_g) = \Lg(d_g)$, while at short distances the dust SG is $\eta$ times stronger than gas SG and we get $\Ld(d_g=0)= \eta L_0$.
This latter equality is important because if (1) the gas SLFC is used instead of the dust SLFC and (2) if a CSL is used instead of a SVSL, the dust SG is underestimated by a factor $\sim 100 \, \eta$ at short distances.
These aspects are of primary interest in an astrophysical context since usually $\eta \gtrsim 10$ which could have important implications for planet formation theories as will be discussed in Sect. \ref{Sec: planet formation consequences}.

\subsection{Crossed contributions}

In this Sect. we take \textbf{a}=d and \textbf{b}=g.
We highlight that the SGFC is commutative with respect to phases \textbf{a} and \textbf{b}.
This could be demonstrated thanks to Newton's third law or by analytical arguments (Appendix \ref{app: commutativity of SGFC}).
Therefore, in the following, we will adopt the notation $\Ldg$ to refer to the dust-gas, or gas-dust, SGFC.
Within this clarification and Eq. \ref{Eq: SGFC general} the dust-gas SGFC is:
\begin{equation}\label{Eq: force correction dust-gas}
\Ldg(d_g, \eta)= \displaystyle \frac{1}{2} d_g^2
             \iint\limits_{u,v=-\infty}^{\infty} 
             \frac{e^{-\frac{u^2}{2}} e^{-\frac{v^2}{2}}}{\left[ d_g^2+(u-v/\eta)^2 \right]^{3/2}}
             \, du \, dv
\end{equation}
Motivated by the SL formalism, we again aim to approach the above double integral by a dust-gas SLFC, but to our knowledge this has never been done before.
In next Sects. we explore for first time what constraints should be satisfied in order to construct a consistent SL which allows to accurately approach the exact dust-gas SGFC.

\subsubsection{Limiting cases for weak and strong layering}\label{sec: eta asymptotic}

When layering is small ($\eta=1$) gas and dust are fully mixed and we immediately retrieve the case of a pure gas disc:
\begin{equation}
\lim\limits_{\eta \rightarrow 1} \Ldg(d_g, \eta)
                    = \Lg(d_g)
\end{equation}
On the other hand, when layering is strong ($\eta>>1$) the dust layer is infinitely thin and we get:
\begin{equation}\label{Eq: planet-disc force correction}
\lim\limits_{\eta \rightarrow +\infty} \Ldg(d_g, \eta)
                     = \sqrt{2 \pi } \, I_p(d_g^2/4) = L^p(d_g)
\end{equation}
where $I_p(x)=x e^x [K_1(x)-K_0(x)]$ and $K_1$ and $K_0$ are the modified Bessel functions of second kind.
Interestingly, $I_p(d_g^2/4)$ is a force function in the limiting case of a planet interacting with a gas disc as exposed in \citep[Sect. 4]{muller_kley_2012}.
Accordingly, we recall the quantity $L^p$ the planet-disc force correction (PDFC).
The PDFC can be approached in the SL formalism and using the same analytical approach conducted in Sect. \ref{subsec:gas and gas} and the results in \citep[Sect. 4.1]{muller_kley_2012} (see Appendix \ref{app: gas SL derivation}), we find the planet-disc SVSL:
\begin{equation}\label{Eq: planet-disc SVSL}
\epsilon_p(d_g) = H_g(\vr) \left[ 1 - \exp\left(-\epsilon_{0,p} \, d_g^{2/3}- \beta d_g^q\right) \right]
\end{equation}
where $\epsilon_{0,p} = \left( \pi/2 \right)^{1/6}$ and the power $2/3$ were obtained by analytical means while $(\beta,q)$ by curve-fitting.
The latter parameters are gathered in Table \ref{tab: fitting parameters}.
It is interesting to note that, contrary to our initial guess, for strong dust layering the dust-gas SLFC tends towards the PDFC rather than towards the dust SLFC.

In summary, the dust-gas SL equals the gas SL or planet-disc SL when the gas-to-dust height, $\eta$, tends towards unity or infinity, respectively.

\subsubsection{Behaviours at short and long separation}\label{sec: asymptotic cg}

\begin{figure}
\centering
\includegraphics[width=\hsize]{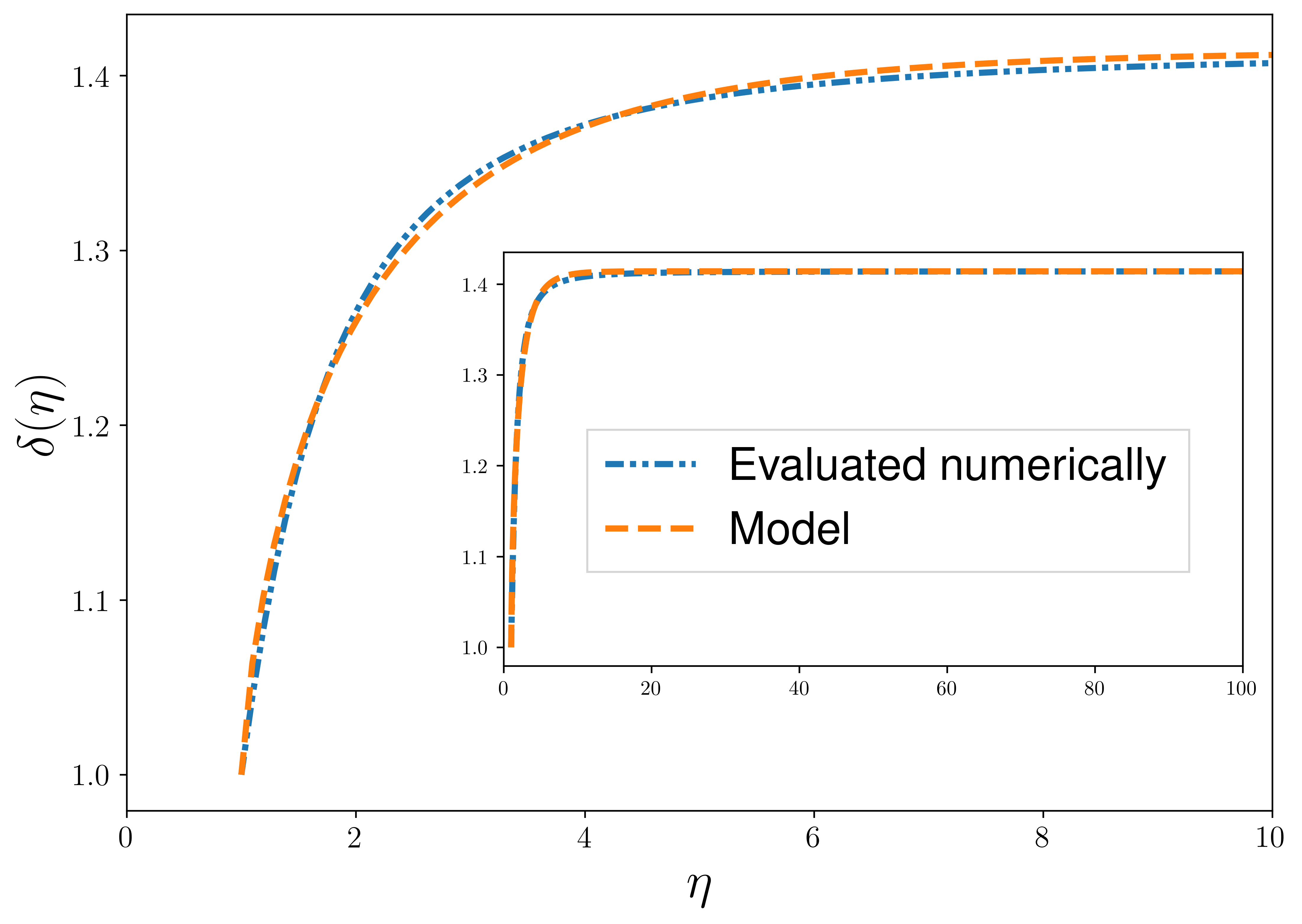}
\includegraphics[width=\hsize]{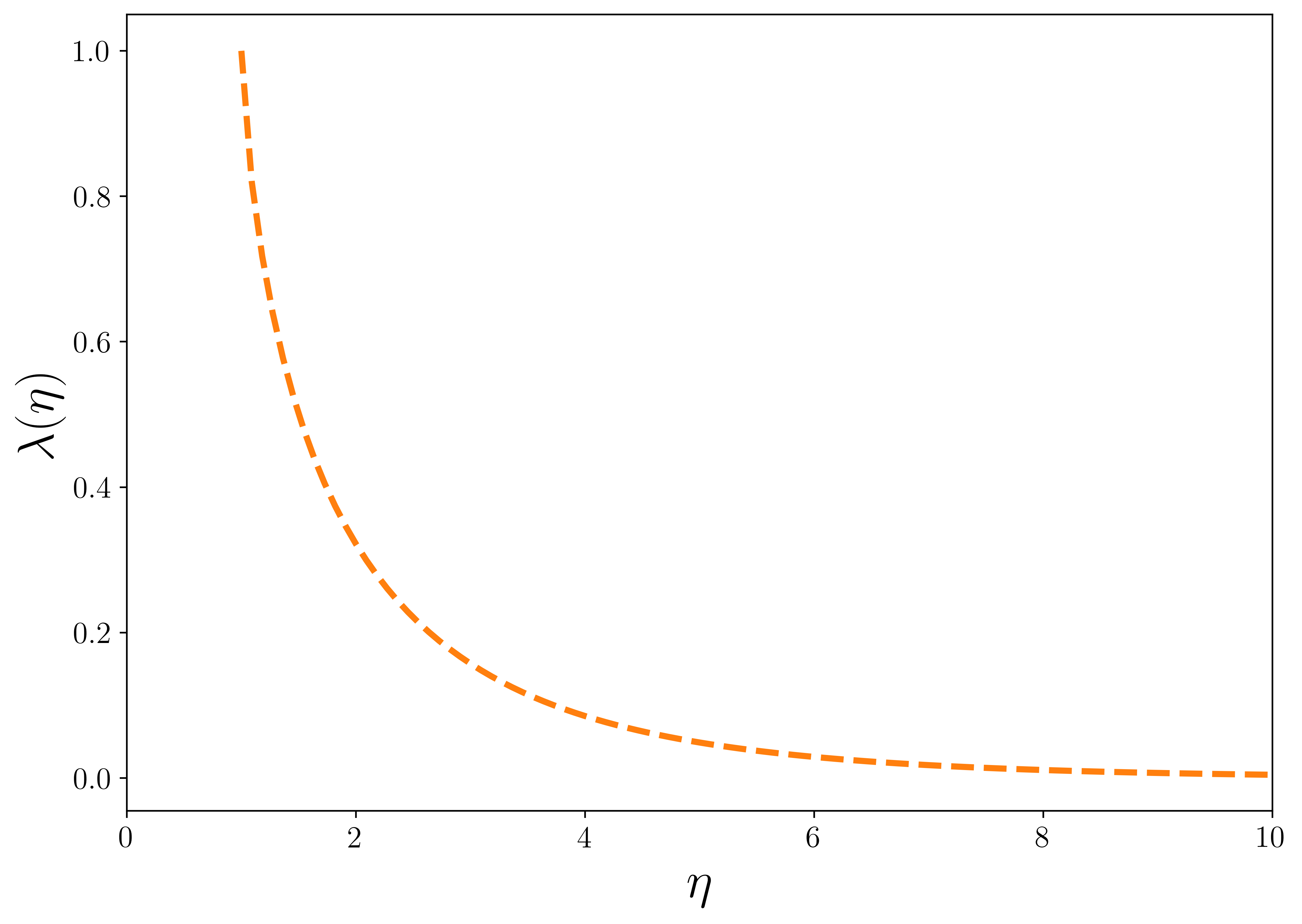}
\caption{\textbf{Model functions for defining the dust-gas SVSL}                                                \\
\emph{Top:} $\delta$ function: model and numerical estimation with respect to the gas-to-dust height ratio ($\eta$). \\
\emph{Bottom:} $\lambda$ model with respect to $\eta$. }
\label{fig: delta, lambda and epsilon_dg}
\end{figure}

\begin{figure}
\centering
\includegraphics[width=\hsize]{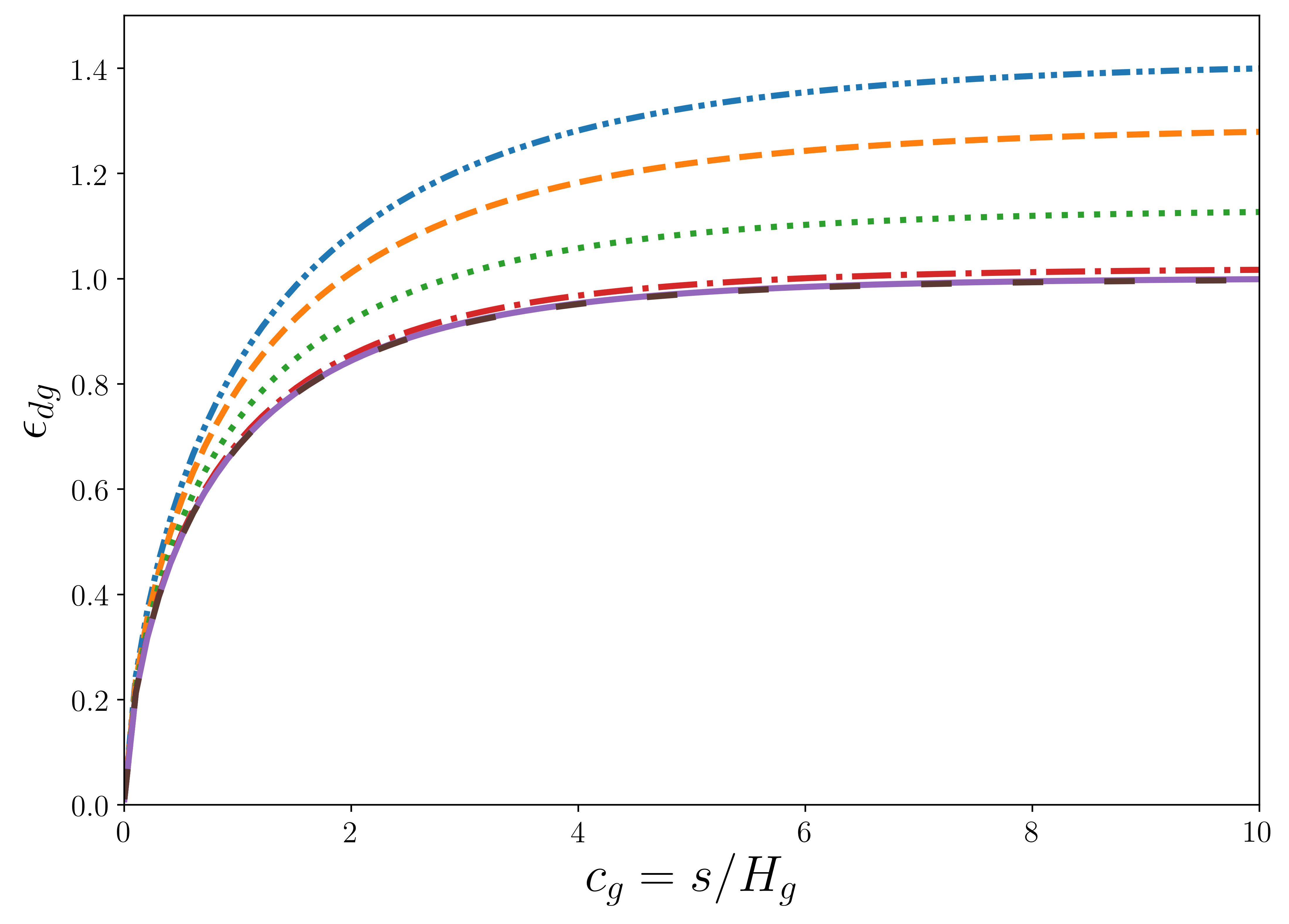}
\includegraphics[width=\hsize]{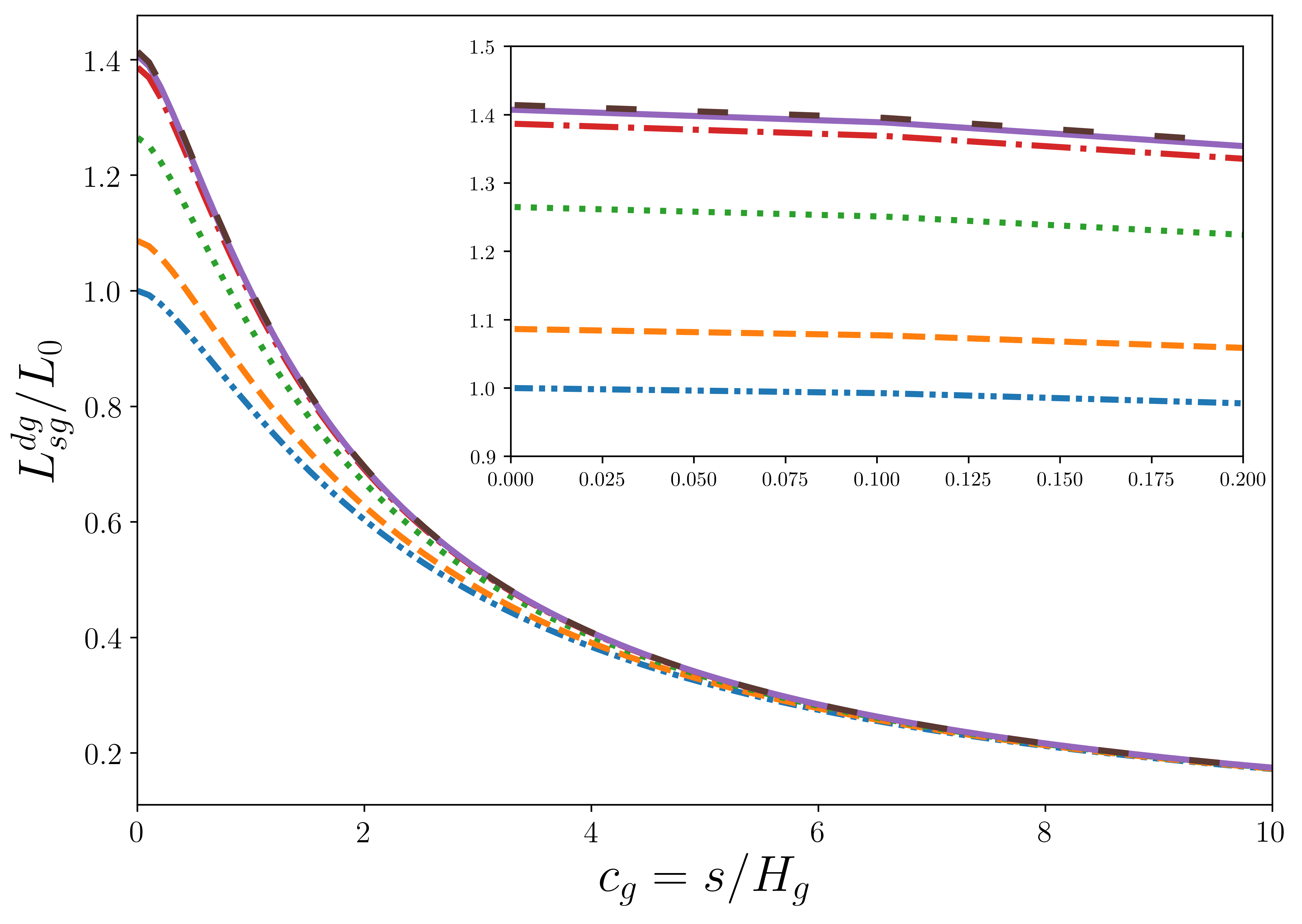}
\includegraphics[width=\hsize]{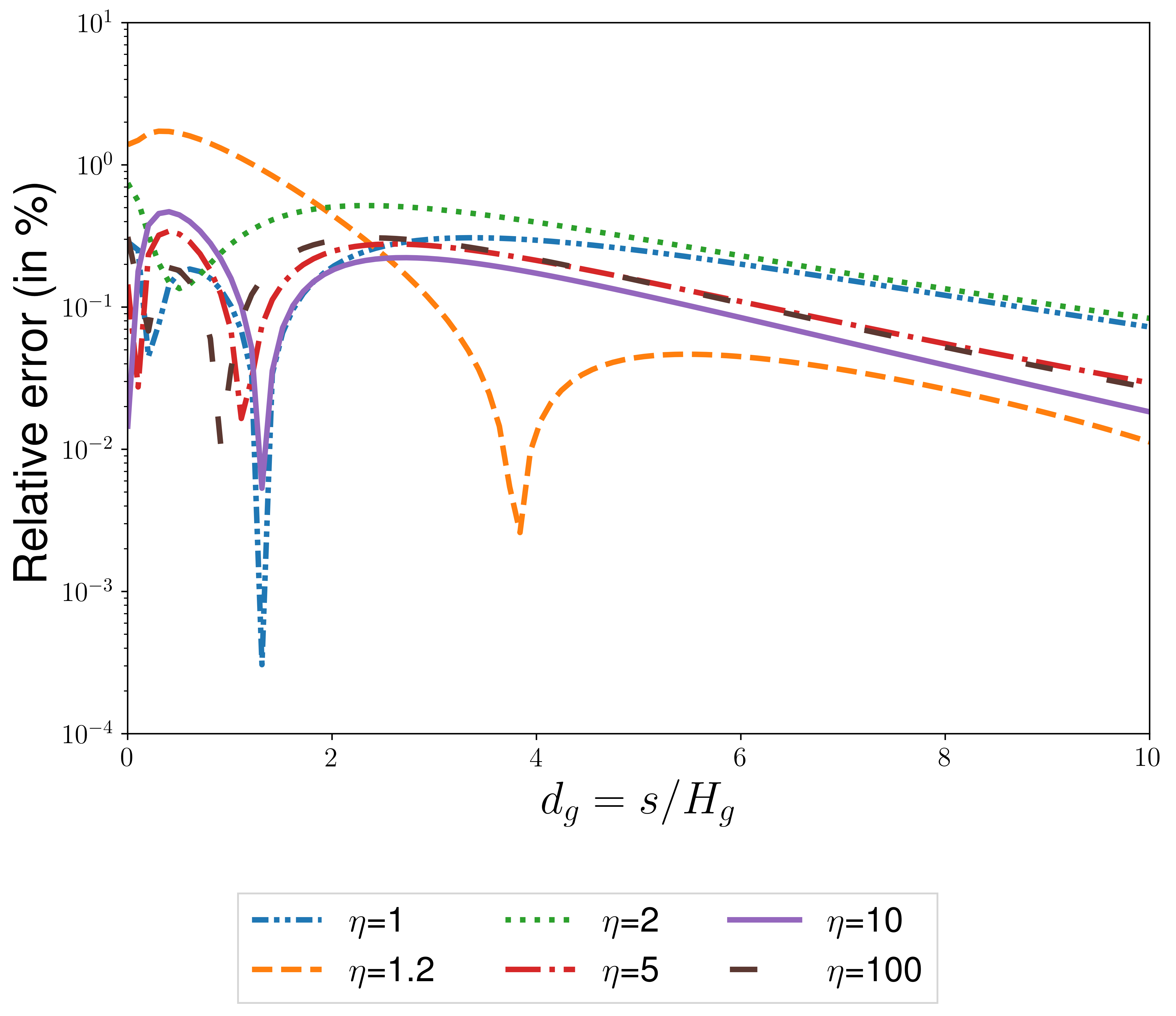}
\caption{\textbf{SG and SL force corrections for dust-gas and space varying smoothing-length}       \\
\emph{Top:} Dust-gas SVSL for different values of $\eta$ (gas-to-dust heights ratio).               \\
\emph{Middle:} Dust-gas SGFC for the same values of $\eta$                                          \\
\emph{Bottom:} Relative error between SGFC and SLFC (in \%): $100\cdot|\mbox{SLFC}/\mbox{SGFC}-1|$. \\
For thin ($\eta>>1$) and thick ($\eta=1$) dust layers the SVSL tends towards the one of the gas and of the planet, respectively.
At short distances the dust-gas SLFC is restricted to the interval $[L_0, \sqrt{2} L_0]$.
The accuracy of the SL method for the gravitational interaction of the gas disc with the embedded dust is better than 2\% for the whole separation range.}
\label{fig: Ldg}
\end{figure}

To study dust-gas SGFC at short distances it is convenient to define:
\begin{equation}\label{Eq: def delta}
\delta(\eta)=\lim\limits_{d_g \rightarrow 0} \Ldg (d_g, \eta)/ L_0
\end{equation}
We did not succeed to express this function in terms of standard mathematical functions but we estimated it numerically in top panel of Fig. \ref{fig: delta, lambda and epsilon_dg} (blue solid line).
As expected, $\delta$ is equal to 1 for $\eta=1$ ($\Ldg$ matches $\Lg$) while for an infinitely thin layer of dust this quantity tends to $\sqrt{2}$. 
Accounting for these boundary conditions and a noticeable curve shape, we get:
\begin{equation}
\delta(\eta) = \sqrt{2} + (1-\sqrt{2}) \, \exp\left[-\gamma (\eta-1)^m\right]
\end{equation}
where the couple $(\gamma, m)$ can be found in Table \ref{tab: fitting parameters}.
This analytic model is also shown in top panel of Fig. \ref{fig: delta, lambda and epsilon_dg} (orange dashed line).
In contrast, for long separations the dust-gas SGFC should satisfy:
\begin{equation}
\lim\limits_{d_g \rightarrow +\infty} \Ldg (d_g, \eta)= \frac{\pi}{d_g}
\end{equation}
This implies that at long distances the dust-gas SVSL should be negligible compared to the square of the distance, $\epsilon_{dg}(d_g)/H_g(r) = o(d_g^2)$, which is undoubtedly satisfied if the SVSL is constant at long distances.

\subsubsection{Dust-gas space varying smoothing length construction}

We have in hand  all necessary information needed to build a consistent dust-gas SVSL. 
We made the choice to look for the dust-gas SVSL under the form of a linear combination of previous asymptotic cases:
\begin{equation}\label{Eq: dust-gas SVSL}
\edg(d_g, \eta)=\lambda(\eta) \, \eg(d_g) + (1-\lambda(\eta)) \, \ep(d_g)
\end{equation}
where $\lambda(\eta) \in [0,1]$.
For matching Sect. \ref{sec: eta asymptotic} constraints, we choose:
\begin{equation}
\begin{array}{ccccc}
\lim\limits_{\eta \rightarrow 1} \lambda(\eta) = 1 &
    \Longrightarrow                          &
    \lim\limits_{\eta \rightarrow 1} \edg(d_g, \eta)       
                   & = & \eg(d_g) \\
\lim\limits_{\eta \rightarrow +\infty} \lambda(\eta) = 0 &    
    \Longrightarrow                          &
    \lim\limits_{\eta \rightarrow +\infty} \edg(d_g, \eta) 
                   & = & \ep(d_g)
\end{array}
\end{equation}
Additionally, from Sect. \ref{sec: asymptotic cg} results and a Taylor expansion (Appendix \ref{app: lambda calculation}), we get:
\begin{equation}\label{Eq: lambda}
\lambda(\eta) = \displaystyle \frac{\epsilon_{0,g}\left( {1}/{\delta(\eta)} \right)^{1/3} - \epsilon_{0,p} }{ \epsilon_{0,g}  - \epsilon_{0,p}}
\end{equation}
This $\lambda$ function is plotted in the bottom panel of Fig. \ref{fig: delta, lambda and epsilon_dg} where we checked that the above mentioned boundary conditions are met.
We also plotted in top panel of Fig. \ref{fig: Ldg} the dust-gas SVSL, defined in Eq. \ref{Eq: dust-gas SVSL}, for different gas-to-dust scale height ratios.
For $\eta=1$ the dust-gas SVSL matches the gas SVSL while for modest dust layering, $\eta\geq5$, the SVSL tends rapidly to the one of the planet-disc interaction case.
This permits to use the approximation: $\edg(d_g) \simeq \ep(d_g)$ for $\eta\geq 5$.
Finally, we found that our results are mathematically consistent provided that $L_0 = \sqrt{\pi}$.
Comparing this theoretical prediction with the value of $L_0$ obtained numerically, we find that both results match with an accuracy of up to six decimals.

\subsubsection{Summary for the dust-gas SVSL}

In middle and bottom panels of Fig. \ref{fig: Ldg} are shown the exact dust-gas SGFC and the error between both estimations for different $\eta$ values, respectively.
The dust-gas SLFC and SGFC curves overlap for the whole $\eta$ and distances range, so we did not plot these two quantities in the same figure to avoid duplication.
This overlap is also reflected in the error, which this time depends on the dust layering: for $\eta \in [1, 5[$ the error is lower than 2 \% and for $\eta \in [5, 100]$ the error is lower than 0.5 \% for the whole distances range.
Compared to both cases studied in Sects. \ref{subsec:gas and gas} and \ref{sec:dust and dust} the error is slightly higher but it remains very acceptable.
We want to stress that the dust-gas SLFC tends rapidly, with respect to $\eta$, to the planet-disc SLFC which makes possible the simplification: 
\begin{equation}
\Ldg(d_g,\eta) = L^p(d_g) \quad \mbox{ or } \quad \edg(d_g, \eta)=\epsilon_p(d_g) \quad \mbox{ if } \quad \eta \geq 5
\end{equation}
This approximation could simplify the numerical treatment.

\section{Numerical treatment}\label{sec: numerical treatment}

The main goal of our study is to implement an accurate SG computation method for multi-fluids in 2D numerical codes.
This could be beneficial for the 2D version  $(r,\theta)$ of hydrodynamical codes such as \textit{RoSSBi3D} \citep{rossbi3d}, \textit{FARGO} \cite{masset_2000} or \textit{Athena} \citep{athena_2008}.
We start by treating the singularity for vanishing separations, responsible of numerical divergences, then we explain explicitly under which conditions results of Sect. \ref{sec: mutual interactions} could be used for estimating SG thanks to FFT methods.
Finally, we quantify computational costs for 2D, N-fluid simulations with SG.

\subsection{Removing numerical divergences}

\begin{figure}
\centering
\includegraphics[width=\hsize]{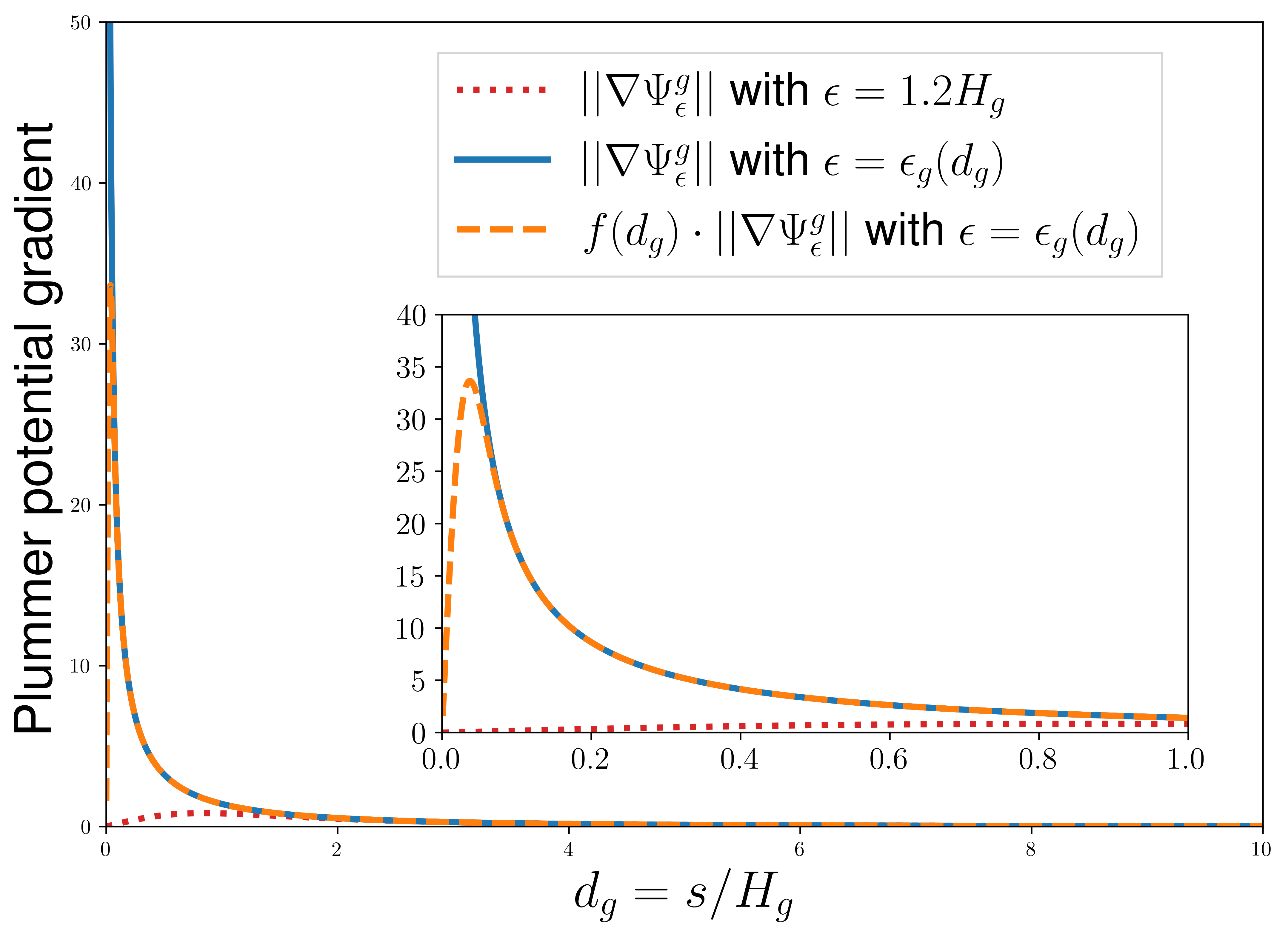}
\caption{\textbf{Normalised Plummer potential gradient with, and without, a tapering function.}\\
Under the SVSL method, the introduction of a tapering function (Eq. \ref{Eq: tapering function}) avoids the divergence of the Plummer potential.
At the same time, the low tapering length, $d_s(= 3/127\simeq 0.024)$, ensures that most of the SG contribution is not lost over short distances ($d_g\leq1.5$). 
Under the CSL approach (red dotted), the gradient cancels at the singularity but most of the short distance contribution to the SG is lost. }
\label{fig: plummer potential gradient}
\end{figure}

From Eq. \ref{Eq: force correction SL} it is obvious that $||\nabla \Psi_\epsilon^{ab}|| \propto d_g^{-1}$ at short separations which makes necessary a product with a tapering function so as to cancel SG for vanishing separations.
The tapering function, $f$, must be equivalent to $d_g^p$ at short distances with $p>1$.
On the other hand it should be equal to unity at large distances.
In order to not loose the accuracy reached in Sect. \ref{sec: mutual interactions} the tapering length should be approximately as large as the finest numerical resolution in the grid: $r_s \sim 3 \min(\Delta r, r \Delta \theta)$ \footnote{In cylindrical coordinates.} \footnote{The factor 3 was chosen for ensuring a safety margin.} where $\Delta r$ and $\Delta \theta$ are the resolution in the radial and azimutal directions, respectively.
For satisfying above constraints we chose arbitrarily $p=2$ and the tapering function as following:
\begin{equation}\label{Eq: tapering function}
f(d_g) =  1-\exp\left[-\frac{1}{2}\left( \frac{d_g}{d_s} \right)^2\right] \isEquivTo{d_g\rightarrow 0} \frac{1}{2} \left( \frac{d_g}{d_s} \right)^2
\end{equation}
where $d_s=r_s/H_g$ is the normalised tapering length.
In Fig. \ref{fig: plummer potential gradient} we show the normalised gradient of the gas Plummer potential with (orange dashed line), and without (blue solid line), the tapering function correction for $d_s=3/127$.
We also plotted same quantity under a CSL assumption (red dotted line).
Such tapering length choice was motivated by the high resolution reached by \citet{rendon_2022}, i.e 146 cells/$H_g$ and 127 cells/$H_g$ in the radial and azimuthal directions.
As expected both Plummer potential gradients have same behaviour for $d_g\gtrsim d_s$ but only the corrected one converges towards $0$ for vanishing distances which should avoid numerical divergences.
Regarding the CSL assumption, we clearly see that SG is underestimated for $d_g\leq 1.5$ but there is no need to resort to a tapering function since the potential gradient cancels analytically at the singularity.
This last statement may be the reason why \citet{muller_kley_2012} did not mention that there were $100\%$ errors with the CSL method at short separations.

\subsection{Numerical calculation with FFT methods}\label{sec: Numerical calculation with FFT methods}

An important question regarding our results is to verify if FFT methods can still be used when we resort to a SVSL.
Indeed, the classic SG computation in 2D was based on a CSL.
We find instructive to remind here from ground how SG is computed, but this time, including a SVSL and a tapering function.
Guided by \citet{2008_baruteau} and \citet[in french]{survi2013} we get the radial component of the SG mass force (divided by $\Sigma_b$) \footnote{We took $G=1$.}:
\begin{equation}\label{eq:radial SG force init}
\begin{array}{ll}
- \displaystyle \frac{\vec{\nabla} \Phi_{\epsilon}^{ab} \cdot \er}{\pi} 
          & \\ [6pt]
\quad =\displaystyle  \iint\limits_{disc}  
             f(d_g) \frac{\Sigma_a(\vr')(\vr-\vr')}{\left[s^2+\epsilon_{ab}(d_g, \vr)^2\right]^{\frac{3}{2}}} 
             \vec{e}_s \cdot \er r' dr' d\theta' & \\ [10pt]
\quad  =\displaystyle  \iint\limits_{disc}  
             f(d_g) \frac{\Sigma_a(\vr') \left( \left(\frac{r}{r'}\right)- \cos{(\theta-\theta')}\right)
             \frac{dr'}{r'} d\theta'  }
             {\left[1+\left(\frac{r}{r'}\right)^2-2\left(\frac{r}{r'}\right)\cos{(\theta-\theta')} + \delta\epsilon_{ab}^2\right]^{\frac{3}{2}}} & 
\end{array}
\end{equation}
where:
\begin{equation}
\begin{array}{lllll}
\delta \epsilon_{ab} &=& \displaystyle \frac{\epsilon_{ab}(d_g, \vr)}{r'} 
                     &=& \displaystyle \frac{r}{r'} h_g(r) \frac{\epsilon_{ab}(d_g, \vr)}{H_g(r)}
\end{array}
\end{equation}
\begin{equation}
d_g  = \displaystyle \frac{r'}{r} \frac{1}{h_g(\vr)} \sqrt{1+\left(\frac{r}{r'}\right)^2-2\left(\frac{r}{r'}\right)\cos{(\theta-\theta')}} 
\end{equation}
and $h_g(\vr)=H_g(\vr)/r$ is the gas disc aspect ratio.
We aim to write the integral, defined by Eq. \ref{eq:radial SG force init}, as a convolution product which is only possible provided that the tapering function, $f$, and the ratio $\delta \epsilon_{ab}$ could be written as $\frac{r}{r'}$ and $\theta-\theta'$ functions.
This constraint is satisfied if the disc aspect ratio is a spatial constant, $h_a(\vr)=h_{0,a}$, in the whole simulation box.
Within this condition the radial component of the SG force is:
\begin{equation}\label{eq:radial SG force}
\begin{array}{ll}
\displaystyle -\frac{\vec{\nabla} \Phi_{\epsilon}^{ab} \cdot \er}{\pi} 
          & \\
\quad \quad = \displaystyle \iint\limits_{disc} 
              \Sigma_a(X',\theta') \, \mathcal{G}_r^{ab}(X-X',\theta-\theta') \, dX' d\theta' & 
\end{array}
\end{equation}
where:
\begin{equation*}
\begin{array}{ll}
\mathcal{G}_r^{ab}(X-X',\theta-\theta') & \\
\quad  =  \displaystyle\frac{f(X-X', \theta-\theta') \left[e^{(X-X')}-\cos{(\theta-\theta')}\right]}
            {\left[1+e^{2(X-X')}-2e^{(X-X')}\cos{(\theta-\theta')} + \delta \epsilon_{ab}(X-X', \theta-\theta')^2 \right]^{\frac{3}{2}}} & \\
\end{array}
\end{equation*}
is the modified radial Green function where we performed the variable substitution: $r=e^X$ and $dr=e^X \, dX$.
For the azimutal component the calculation is similar and we obtain:
\begin{equation}\label{eq:azimuth SG force}
\begin{array}{ll}
\displaystyle-\frac{\vec{\nabla} \Phi_{\epsilon}^{ab} \cdot \et}{\pi}
            &= \displaystyle \iint\limits_{disc} 
            \Sigma_a(X',\theta')\, \mathcal{G}^{ab}_\theta(X-X',\theta-\theta') \, dX' d\theta'
\end{array}
\end{equation}
where:
\begin{equation*}
\begin{array}{ll}
\mathcal{G}^{ab}_\theta(X-X',\theta-\theta') & \\ 
\quad   = \displaystyle \frac{f(X-X',\theta-\theta') \, \sin{(\theta-\theta')}}
          {\left[1+e^{2(X-X')}-2e^{(X-X')}\cos{(\theta-\theta')} + \delta\epsilon_{ab}(X-X', \theta-\theta')^2 \right]^{\frac{3}{2}}}    & \\
\end{array}
\end{equation*}
is the modified azimutal Green function.
The rewriting of both integrals, defined by Eqs. \ref{eq:radial SG force} and \ref{eq:azimuth SG force}, permits the last writing in terms of Fourier transforms:
\begin{equation}\label{Eq: FFTW}
\begin{array}{ccc}
\vec{f}_{2D,\epsilon}^{a\rightarrow b}(\vr)\cdot \er  
                                    &=& \Sigma_b(\vr) \,
                                        \fourier^{-1}
                                        \left[ 
                                        \fourier\left(\Sigma_a\right) \ast
                                        \fourier\left(\mathcal{G}^{ab}_r\right)
                                        \right] \\ [6pt]
\vec{f}_{2D,\epsilon}^{a\rightarrow b}(\vr)\cdot \et  
                                    &=& \Sigma_b(\vr) \,
                                        \fourier^{-1}
                                        \left[ 
                                        \fourier\left(\Sigma_a\right) \ast
                                        \fourier\left(\mathcal{G}^{ab}_\theta\right)
                                        \right] 
\end{array}
\end{equation}
where $\fourier$ and $\fourier^{-1}$ are the Fourier transform operator and its inverse.
The symbol $\ast$ denotes the convolution operator.
In practice such quantities are computed numerically thanks to fast Fourier modules which highly accelerate numerical computation.
For instance, in the 2D version of \rossbi{} \citep{rossbi3d} this computation is made possible thanks to the FFTW3 library \citep{FFTW05}.
It is important to highlight that the use of Fourier transforms is made possible if \textbf{(1)} a logarithmic mesh is used in the radial direction.
This condition stems from the radial variable substitution which enabled us to obtain the formulation with the modified Green's functions.
Furthermore, \textbf{(2)} the function $\Sigma_a$ must be periodic in radial and azimutal directions in order to be able to use Fourier transforms.
Periodicity is obviously satisfied in the azimutal direction but in the radial direction this is not necessarily the case.
Such periodicity is artificially ensured thanks to a zero-padding: the radial domain is doubled and the function $\Sigma_a$ is set to 0 in half of the domain.
See \citet[Fig. III.10][in french]{survi2013} for an example.
Finally, it is necessary that \textbf{(3)} the disc aspect ratio of gas and dust must be constant with respect to $r$ (but it could be a time varying function).
If such condition is not satisfied the formulation as a convolution product is not possible.

We take the opportunity to clarify that the aforementioned assumption of a constant disc aspect ratio used in this Sect. only permits to resort to FFT methods for accelerating numerical calculations.
This condition is not new since it was already implicit in the classical calculation of the SG by FFT.
For general scale heights, direct summation in the radial direction and Fourier transforms in the azimuth direction, ensured by the periodicity, are a straightforward solution.

\subsection{Computational costs}

For SG simulations the computational endeavour is a non-negligible aspect and particularly for high-resolution simulations.
For conciseness and without loss of generality we will only treat the computation of SG in the radial direction and we don't account the Fourier transforms of the modified Green's functions since they are computed only once.
For the case of a bi-fluid simulation next Fourier transforms should be performed at each time step:

\noindent \textbf{Gas into gas}
\begin{equation}
\fourier (\Sigma_g) \quad\mbox{and} \quad 
                          \fourier^{-1}
                          \left[ 
                          \fourier\left(\Sigma_g\right) \ast
                          \fourier\left(\mathcal{G}^{g}_r\right)
                          \right] 
\end{equation}
\textbf{Dust into dust}
\begin{equation}
\fourier (\Sigma_d)  \quad\mbox{and} \quad 
                          \fourier^{-1}
                          \left[ 
                          \fourier\left(\Sigma_d\right) \ast
                          \fourier\left(\mathcal{G}^{d}_r\right)
                          \right] 
\end{equation}
\textbf{Dust into gas}
\begin{equation}
\fourier (\Sigma_d)  \quad\mbox{and} \quad 
                          \fourier^{-1}
                          \left[ 
                          \fourier\left(\Sigma_d\right) \ast
                          \fourier\left(\mathcal{G}^{dg}_r\right)
                          \right] 
\end{equation}
\textbf{Gas into dust}
\begin{equation}
\fourier (\Sigma_g)  \quad\mbox{and} \quad 
                          \fourier^{-1}
                          \left[ 
                          \fourier\left(\Sigma_g\right) \ast
                          \fourier\left(\mathcal{G}^{dg}_r\right)
                          \right] 
\end{equation}

\noindent From above recapitulation we observe that the Fourier transforms of gas and dust densities are duplicated which reduces the total number of Fourier operations to 6 during a numerical treatment.
Therefore, the numerical endeavour for computing SG for a bi-fluid is 3 times larger with respect to the case of a single fluid.
In general, for $N$ fluids with different scale heights and interacting through SG the amount of Fourier transforms is:
\begin{equation}
N^2+N = \displaystyle \underbrace{\left( 2 N^2 \right)}_{\mbox{All $N$-tuples  combinations}}-\underbrace{\left(N^2-N \right) }_{\mbox{Duplicates}}
\end{equation}
Compared to standard self-gravitating simulations on a unique gaseous phase, the computation time is crudely multiplied by a factor $\sim N^2/2$ for large $N$.
Of course this amount can be decreased assuming that some of the involved fluids have same scale heights.

%

\section{Discussion}\label{sec:discussion}

In this Sect. we treat the possible impact of our findings regarding planet migration and the early stage of planetary formation.
Then we identify the limitations of the SVSL approach due to our initial assumptions and to the specificity of the studied problem. 
Finally, we propose possible ways to improve and test our model.  

\subsection{Consequences for planet-disc interaction}\label{Sec: planet disc interaction}

It is well known that planets can migrate due to tidal interactions with the gas disc.
This is the case of type I migration in which the planet can exchange angular momentum with Lindblad and co-rotation resonances \citep[for a review]{baruteau_migration}.
In 2D numerical simulations, the value $\epsilon_p/H_g(\vr)=0.3-0.6$ is often used to match the total torque exerted on a planet in 3D simulations \citep{masset_2002, tanaka_2002}.

Similarly, to our results of Sect. \ref{subsec:gas and gas}, the planet-disc SLFC is not well captured by a CSL for separations inferior to $\sim 1.5 H_g$.
Although we do not question the agreement between the 2D and 3D simulations of the planet-disc interaction, we do believe that our SVSL may be better suited than an adjustment factor.
Therefore, it might be constructive to verify whether \textbf{(1)} the results of 3D simulations can be retrieved using our SVSL and check whether \textbf{(2)} the widely used $\epsilon_p/H_g(\vr)=0.3-0.6$ factor could be retrieved from an analytical basis stemming from our SVSL.
We also think that our SVSL approach could be helpful for addressing 2D simulations of low-mass planets embedded in a self-gravitating disc.
This would require, though, to make some improvements: the vertical layering due to SG must be accounted, as discussed in Sect. \ref{sec: stratification, asymmetries and disc time evolution}, and the vertical stratification of the gas disc due to the planet gravitation must be assessed, as noted by \citet{muller_kley_2012}.
Both aspects raised in this paragraph require a more detailed work which is out of the scope of present paper.

\subsection{Consequences for planet formation theories}\label{Sec: planet formation consequences}

One attractive planet formation scenario is based on the persistence of gaseous vortices in PPDs.
Its main interest is in the strong capture efficiency of the vortices and their ability to confine large concentrations of dust-grains that could collapse to form planetesimals or a planetary core \citep{barge_1995}.
Even if observational findings seem encouraging in this way \citep{varga_2021}, numerical simulations have not yet concluded that vortices could form objects bond by gravity.
The results of the present paper offer the possibility to carry out new numerical simulations that correctly account for SG in the gas and dust components of PPDs. 
Particularly, we expect that these new simulations could answer the questions raised by the vortices.
Indeed, as demonstrated in Sect. \ref{sec:dust and dust}, dust SG could be underestimated by a factor $\sim 2 000$ for $\eta=20$ at short separations.
At the same time the estimation of the dust-gas SLFC could permit to quantify with an acceptable accuracy a possible gaseous envelope capture by dust clumps.
From theoretical analysis it was found that SG inhibit vortices formation by Rossby Wave Instability \citep{lovelace_2013} which was later confirmed by numerical simulations \citep{baruteau_2016, regaly_vorobyov_2017,tarczay_2021}.
In the latest numerical work to date, \citet{rendon_2022} also found that vortices cannot survive in massive PPD and they provided a stability criterion that vortices should satisfy in order to resist SG destabilising effects.
At the light of our findings, previous simulations results should be checked anew for understanding at which extent the SVSL affects theoretical predictions on vortices survival in self-gravitating PPDs.

\subsection{Limitations, improvements and tests}

The limitations of our model are inherent to the initial assumptions we made about the vertical structure of the disc in relation to gas and dust stratification. 

\subsubsection{Stratification and disc evolution}\label{sec: stratification, asymmetries and disc time evolution}

The SGFC studied in this paper is based on the vertical integration of Eq. \ref{Eq: SGFC general} in the particular case of a vertically isothermal disc.
However the vertical structure could be affected by different mechanisms which implies that for any vertical stratification different from the Gaussian stratification, the entire work performed in this paper should be repeated and adapted.
For instance, that's the case when including the disc vertical SG for gas (but neglecting the vertical component of the central object gravity) which modifies the Gaussian distribution into: $\cosh{\left(\frac{z}{Q H_g}\right)}^{-2}$ where $Q$ is the Toomre's parameter \citep{lodato_2007}.
A similar layering should occur for the dusty disc and we also expect that the crossed gravitational interaction between both phases could impact their respective vertical stratification.
Indeed, gas SG could decrease dust scale height by a factor $\sim$ 2 \citep{2021ApJ...909..136B} and we expect that a strong dust layering will also modify gas vertical structure in correlation with the dust-to-gas density ratio.

The global vertical structure of PPDs evolves in time due to accretion heating \citep{2019ApJ...881...56S} and stellar irradiation \citep{2021_wu_lithwick}, amongst others.
Instabilities could also generate time variable structures which could affect, locally, the vertical stratification of the flow.
This was reported, for instance, in 3D vortices simulations \citep{2010_meheut}, for rings and gaps generated by poloidal magnetic winds \citep{2017MNRAS.468.3850S} and for spiral density waves \citep{2018MNRAS.476.5115R}.
The method described in this paper is not limited to a steady vertical stratification, but is compatible with the global and local time evolution of the vertical structure, provided that the time dependence of gas and dust scale heights is known.
This could be done, for instance, with the 2D1D strategy adopted by \citet{2009_crida}.
However, in the particular case of spatially constant aspect ratios, $h_a(\vr,t)=h_a(t)$, the computational benefits from the FFT method could be lost.
Indeed, this would require to perform the Fourier transforms of Green's functions at each time step instead of a unique computation for a steady vertical stratification.
For a simulation with only gas, this would result in 3 Fourier transforms at each time step instead of 2.
We think that for other SG computational methods, the computational costs would be unaffected.

\subsubsection{Layering of the dust particles}

In this paper, the dust component of PPDs is considered as a pressure-less fluid which is sufficiently mixed by the turbulent motions of the gas disc to be maintained in an equatorial sub-layer. 
This assumption requires that the dust particles and gas aerodynamic coupling is governed by a Stokes number less than unity and that dust is adequately diluted in the gas to avoid frequent mutual interactions (if the dust-to-gas mass-ratio $ \lesssim 1$) \citep{2004Garaud}. 
We also assumed that small-scale turbulence is maintained in the gas disc by a mechanism that we disregarded. 
It is interesting to note that outside the bi-fluid pressure-less assumption, turbulent stirring may be replaced by collisional and/or gravitational stirring with the formation of a sub-layer of solid particles whose scale height is different from the one deduced from turbulent stirring in the introduction. 
In such cases the necessary smoothing lengths will be different but the SVSL approach should remain unchanged.

\subsubsection{Additional test}

A relevant test for our results consists in a comparison with the vertically averaged SG obtained thanks to a 3D simulation.
In order to satisfy our assumptions, the 3D disc should be vertically isothermal and the vertical SG should be discarded.
In addition, for the bi-fluid version, the dust layer should be as smooth as possible.
This simulation is out of the scope of current paper but still an interesting lead that the authors want to explore in a near future.

\section{Conclusions}

We revealed contradictions and shortcomings of the CSL method commonly used to compute the contribution of SG in 2D numerical simulations.
In particular, we found that, from short to intermediate separations ($d_g\leq1.5$), the SG force is analytically underestimated with an error that reaches $100\%$ at the singularity. 
We corrected these inconsistencies replacing the CSL, $\epsilon_g=const.$, by a space dependent function, $\epsilon_g=\epsilon_g(d_g)$ (SVSL).
We found that, for a gas disc, the SVSL dependence that better fits the exact SGFC is:
\begin{equation}
\egc=\sqrt{2} \, H_g(\vr) \, \left[1-\exp\left(-\frac{\epsilon_{0,g}}{\sqrt{2}} d_g^{2/3}-\alpha d_g^n\right) \right]
\end{equation}
where $\epsilon_{0,g}=\pi^{1/6}$ and $(\alpha,n)$ can be found in Table \ref{tab: fitting parameters}.
This SVSL approach can be extended to the dust disc using the following dust SL:
\begin{equation}
\ed(d_g, \eta)= \frac{\sqrt{2} \, H_g(\vr)}{\eta} \, \left[1-\exp\left(-\frac{\epsilon_{0,g}}{\sqrt{2}} (\eta d_g)^{2/3}-\alpha (\eta d_g)^n\right) \right]
\end{equation}
where $\eta=\langle H_g\rangle /\langle H_d \rangle $ is the gas-to-dust height ratio.
As a side result we also found the planet-disc SVSL:
\begin{equation}
\epsilon_p(d_g) = H_g(\vr) \left[ 1 - \exp\left(-\epsilon_{0,p} \, d_g^{2/3}- \beta d_g^q\right) \right]
\end{equation}
where $\epsilon_{0,p}=\left(\frac{\pi}{2}\right)^{1/6}$ and $(\beta, q)$ can be found in Table \ref{tab: fitting parameters}.
The crossed gravitational interaction of the gas with the embedded dust can also be evaluated through the SL method.
We constructed this dust-gas SL from a linear combination of aforementioned planet-disc and gas SVSL:
\begin{equation}
\edg(d_g, \eta)=\lambda(\eta) \, \eg(d_g) + (1-\lambda(\eta)) \, \ep(d_g)
\end{equation}
where the analytical expression of $\lambda$ is given in Eq. \ref{Eq: lambda}.
All these new expressions for the SVSL are valid for any scale height if the stratification is Gaussian and remain compatible with the common FFT method for evaluating SG in 2D hydrodynamical simulations provided that the gas and dust disc aspect ratios are constant.
Finally, the use of a tapering function is required to avoid numerical divergences.

The proposed SVSL correction decreases the error up to factors 200 with respect to the latest CSL prescription proposed by \citet{muller_kley_2012}.
In particular our SVSL allows to match the SGFC with a high accuracy, even at the singularity ($d_g\rightarrow0$).
Regarding the dust SLFC, we found that it is proportional to $\eta$ at short separations.
This result combined with the improvement brought by the SVSL method demonstrates that dust SG was generally underestimated by a factor $\sim 100 \, \eta$ at short separations (compared to the CSL only based into the gas SL).

Our planet-disc SVSL could affect the torque exerted by the self-gravitating disc on a planet, which suggests that type I planet migration could be impacted.
We also think that the improvements we have made in the computation of the SG terms will be decisive in the future 2D simulations of PPDs inhabited by a large-scale vortex.
Hence, we forecast a much better description of the evolution of the dust-gas mixture in the core of the vortices with our model than with standard ones; we also expect significant consequences in the simulations of planetesimal construction. 
Indeed, on the one hand dust SLFC could favour gravitational binding of the dust clumps trapped in the vortex and, on the other hand, dust-gas SLFC enables to follow the coupled evolution of dust and gas in the various clumps. 
We speculate that gas could be dragged with dust during the collapse before being trapped in a gaseous envelope around a dusty core.

\begin{acknowledgements}
We thank the referee for her/his helpful comments that enriched the discussion.
We would like to warmly acknowledge Stéphane Le Dizès for fruitfull discussions during the preparation of the paper and financial support during the PhD thesis.
S.R.R. thanks Clément Baruteau for useful discussions and Andrej Hermann for proofreading the article.
Co-funded by the European Union (ERC, Epoch-of-Taurus, 101043302). Views and opinions expressed are however those of the author(s) only and do not necessarily reflect those of the European Union or the European Research Council. Neither the European Union nor the granting authority can be held responsible for them.
\end{acknowledgements}

\bibliographystyle{aa}
\bibliography{bibliography}

\begin{appendix}

\section{Gas and planet SVSL derivation}\label{app: gas SL derivation}

\begin{figure}
\centering
\includegraphics[width=\hsize]{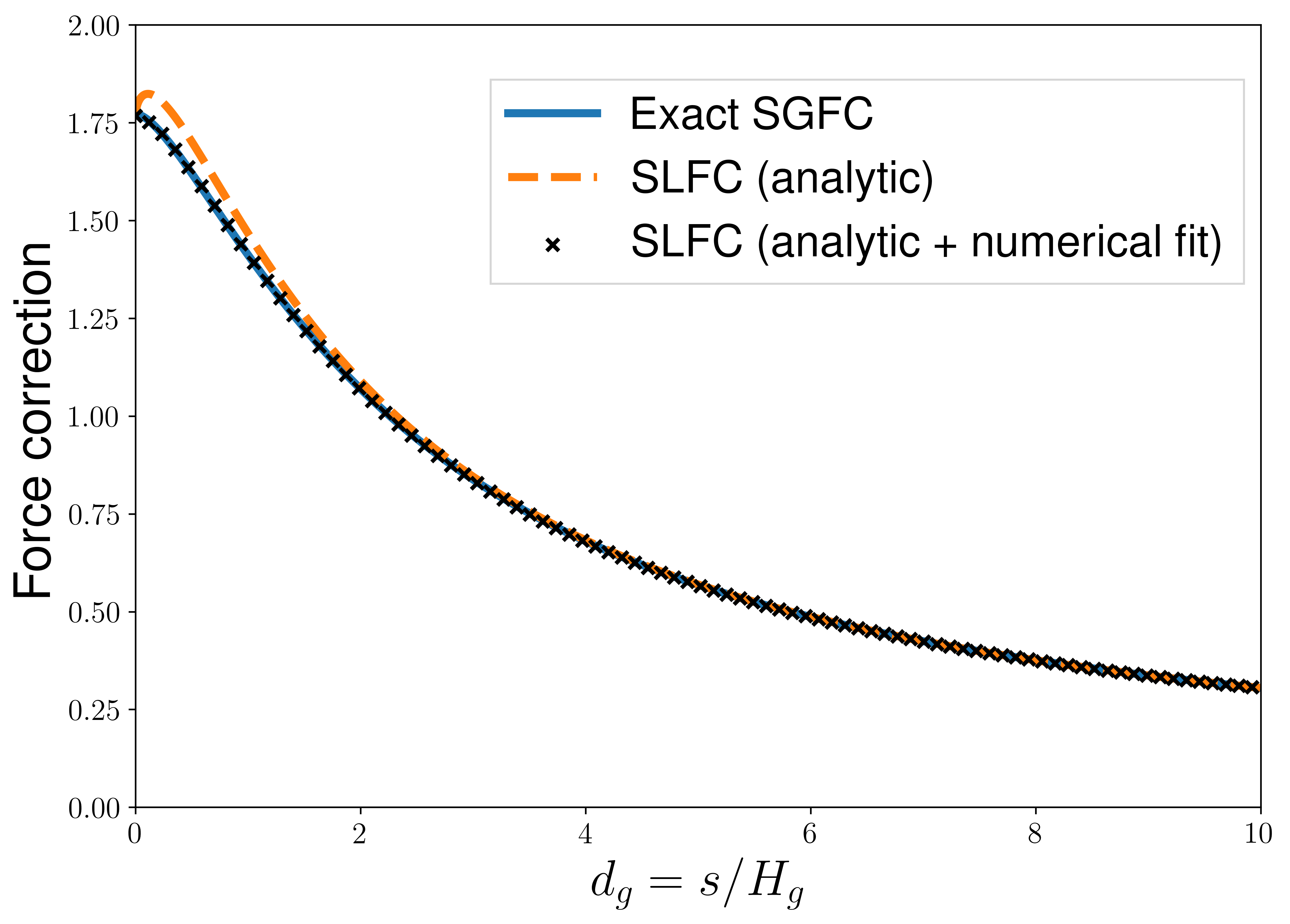}
\includegraphics[width=\hsize]{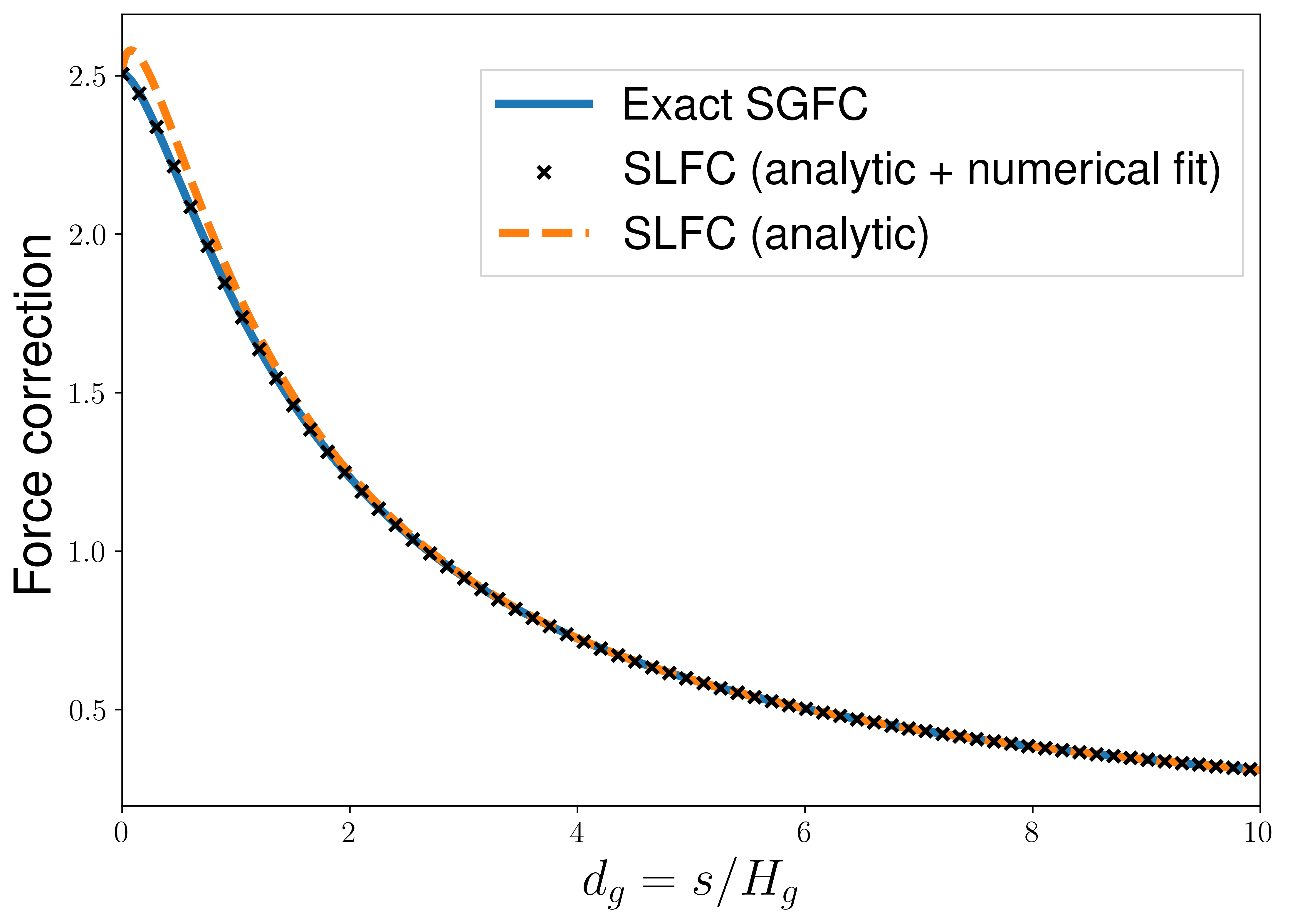}
\caption{\textbf{Gas and planet-disc force corrections}. \\
\emph{Top:} Gas force correction.                        \\
\emph{Bottom:} Planet-disc force correction.             \\
For each force correction we plotted the exact SGFC evaluated numerically (blue solid line), 
the SLFC approximation accounting only the analytical correction (orange dashed line)
or accounting an additional numerical fit (black cross markers).}
\label{fig: Lg and Lp analytic + numerical}
\end{figure}

From \citet{muller_kley_2012} results we chose to look for a SVSL under the form:
\begin{equation}
\epsilon_g(d_g)/H_g(r) = \sqrt{2}\left[1-\exp\left(-\epsilon_{0,g} d_g^k \right) \right] 
\end{equation}
with the constraint $L_\epsilon^g(d_g=0)=L_0$.
From a Taylor expansion we get:
\begin{equation}
L_\epsilon^g(d_g)\isEquivTo{d_g\rightarrow 0} \frac{\pi d_g^2}{\left[ d_g^2+\epsilon_{0,g}^2 d_g^{2 k} \right]^{3/2}} 
\end{equation}
The SLFC converges towards $L_0$ if and only if $k=2/3$ and $\epsilon=[\pi/L_0]^{1/3}$.
This model can be highly improved thanks to an additional term in the exponential under the form $-\alpha d_g^n$.
The values of $(\alpha,n)$ were obtained by numerical fitting to the exact SGFC.
In Fig. \ref{fig: Lg and Lp analytic + numerical} top panel we compare the exact SGFC (blue solid line) with the SVSL obtained only by analytical means (orange dashed line) or the SVSL on which we also introduced the numerical fit (black cross markers).  
We observe that the unique analytical correction decreases the error to less than 5\% for the whole distance range.

Same procedure was used for estimating the SVSL of a planet interacting with a disc.
In Fig. \ref{fig: Lg and Lp analytic + numerical} bottom panel we compare the SVSL obtained only by analytical means (orange dashed line) with the SVSL which also includes the numerical fit (black cross markers). 
Again, the accuracy of the planet-disc SVSL method is excellent compared to the CSL method.

\section{Commutativity of the SGFC for different phases}\label{app: commutativity of SGFC}

Thanks to the principle of action-reaction applied to the whole dusty disc and the whole gas disc \footnote{This rationale also works for an elementary dust volume and an elementary gas volume. But not between an elementary volume of dust and a disc of gas (and vice-versa).} we get:
\begin{equation}
\iint\limits_{disc} \vec{F}_{sg}^{d\rightarrow g}(\vr) d^2\vr 
                 = 
- \iint\limits_{disc} \vec{F}_{sg}^{g\rightarrow d}(\vr') d^2\vr'
\end{equation}
which implies that $L_{sg}^{d\rightarrow g}(d_g, \eta)=L_{sg}^{g\rightarrow d}(d_g, \eta)$.
This commutativity is also easily recovered from Eq. \ref{Eq: SGFC general}:
\begin{equation}
\begin{array}{lll}
L_{sg}^{g\rightarrow d}(d_g,\eta) 
  &=& \displaystyle\frac{1}{2} \frac{d_d^3}{d_g} 
  \iint\limits_{u,v=-\infty}^{\infty}
  \frac{e^{-\frac{u^2}{2}}e^{-\frac{v^2}{2}}}{\left[d_d^2+(u-\eta v)^2\right]^{3/2}} \, du \, dv \\ [20pt]
  & & \quad \mbox{but} \quad d_d = \eta d_g                                                      \\ [5pt]
  &=& \displaystyle\frac{1}{2} d_g^2 
  \iint\limits_{u,v=-\infty}^{\infty}
  \frac{e^{-\frac{u^2}{2}}e^{-\frac{v^2}{2}}}{\left[d_g^2+(u/\eta-v)^2\right]^{3/2}} \, du \, dv \\
  &=& L_{sg}^{d\rightarrow g}(d_g,\eta) 
\end{array}
\end{equation}
As a consequence, in the whole paper we only use the notation $L_{sg}^{dg}$.

\section{Derivation of $\lambda$ and $L_0$}\label{app: lambda calculation}

The Taylor expansion of Eq. \ref{Eq: force correction SL}, where \textbf{a}=d and \textbf{b}=g, in the vicinity of $d_g \sim 0$ is:
\begin{equation}
\begin{array}{lcl}
\Ldg(d_g,\eta) 
  &\isEquivTo{d_g \rightarrow 0} & \displaystyle \frac{\pi}{\left[ \lambda(\eta)(\epsilon_{0,g}-\epsilon_{0,p})+\epsilon_{0,p} \right]^{3}} 
\end{array}
\end{equation}
But the SLFC also satisfies $\lim\limits_{d_g \rightarrow 0} \Ldg (d_g, \eta) = \delta(\eta) L_0$ (Eq. \ref{Eq: def delta}).
The equalisation of both equations leads to:
\begin{equation}
\lambda(\eta) = \displaystyle \frac{\epsilon_{0,g}\left( {1}/{\delta(\eta)} \right)^{1/3} - \epsilon_{0,p} }{ \epsilon_{0,g}  - \epsilon_{0,p}}
\end{equation}
Since $\delta(\eta=1)=1$, it is immediate that $\lambda(\eta=1)=1$.
The constraints of Sect. \ref{sec: eta asymptotic} require that the $\lambda$ function cancels for infinite gas-to-dust scale height ratios.
Considering that $\delta(\eta\rightarrow\infty)=\sqrt{2}$, above condition is only possible if:
\begin{equation}\label{Eq: relation of L0}
\frac{\epsilon_{0,p}}{\epsilon_{0,g}}=\left(\frac{1}{2}\right)^{1/6}
\end{equation}
where $\epsilon_{0,g}=\left(\frac{L_0}{\pi}\right)^{1/3}$ and $\epsilon_{0,p}=\left(\frac{\pi}{2}\right)^{1/6}$.
The mathematical relation \ref{Eq: relation of L0} implies the constraint $L_0=\sqrt{\pi}$.

\end{appendix}

\end{document}